\documentclass[aps, prx, reprint, superscriptaddress, noeprint]{revtex4-2}
\usepackage{amsmath}
\usepackage{graphicx,bm,braket}
\usepackage[colorlinks,citecolor=purple]{hyperref}
\usepackage{placeins}
\usepackage[table]{xcolor}
\usepackage{color}
\usepackage[caption=false]{subfig}
\usepackage{booktabs, threeparttable}
\usepackage{tabularx} 
\usepackage{makecell}
\usepackage{cellspace}

\addparagraphcolumntypes{X}
\newcolumntype{C}{>{\centering\arraybackslash}X}
\newcolumntype{L}{>{\raggedright\arraybackslash}X}
\DeclareMathAlphabet\mathbfcal{OMS}{cmsy}{b}{n}

\newcommand{\stront}{Sr$_2$RuO$_4$ }

\definecolor{inter_orb_singlet_xyxz_xyyz}{HTML}{e41a1c}
\definecolor{inter_orb_singlet_xzyz}{HTML}{e41a1c}

\definecolor{singlet_d_x2_y2}{HTML}{ff7f00}

\definecolor{triplet_xy}{HTML}{377eb8}
\definecolor{triplet_xz_yz}{HTML}{377eb8}
\definecolor{triplet_d_x2_y2}{HTML}{377eb8}

\definecolor{inter_orb_triplet_xyxz_xyyz}{HTML}{984ea3}
\definecolor{inter_orb_triplet_xzyz}{HTML}{984ea3}

\definecolor{singlet_f}{HTML}{4daf4a}

\begin{document}
\title{Inter-orbital singlet pairing in Sr$_2$RuO$_4$: a Hund's superconductor} 
\author{Stefan K\"aser}
\affiliation{Max-Planck-Institut  f\"ur  Festk\"orperforschung, Heisenbergstrasse  1,  70569  Stuttgart,  Germany}
\affiliation{Department of Physics, Friedrich-Alexander Universit\"at Erlangen-N\"urnberg, Germany}
\author{Hugo U. R. Strand}
\affiliation{School of Science and Technology, \"Orebro University, Fakultetsgatan 1, SE-701 82, \"Orebro, Sweden}
\affiliation{Department of Physics, Chalmers University of Technology, SE-412 96 Gothenburg, Sweden}
\affiliation{Center for Computational Quantum Physics, Flatiron institute, Simons Foundation, 162 5th Ave., New York, 10010 NY, USA}
\author{Nils Wentzell}
\affiliation{Center for Computational Quantum Physics, Flatiron institute, Simons Foundation, 162 5th Ave., New York, 10010 NY, USA}
\author{Antoine~Georges}
\affiliation{Coll{\`e}ge de France, 11 place Marcelin Berthelot, 75005 Paris, France}
\affiliation{Center for Computational Quantum Physics, Flatiron institute, Simons Foundation, 162 5th Ave., New York, 10010 NY, USA}
\affiliation{CPHT, CNRS, Ecole Polytechnique, IP Paris, F-91128 Palaiseau, France}
\affiliation{DQMP, Universit{\'e} de Gen{\`e}ve, 24 quai Ernest Ansermet, CH-1211 Gen{\`e}ve, Suisse}
\author{Olivier Parcollet}
\affiliation{Center for Computational Quantum Physics, Flatiron institute, Simons Foundation, 162 5th Ave., New York, 10010 NY, USA}
\affiliation{Universit\'e Paris-Saclay, CNRS, CEA, Institut de Physique Th\'eorique, 91191, Gif-sur-Yvette, France}
\author{Philipp Hansmann}
\affiliation{Max-Planck-Institut  f\"ur  Chemische Physik fester Stoffe, N\"othnitzerstrasse 40,  01187 Dresden,  Germany}
\affiliation{Department of Physics, Friedrich-Alexander Universit\"at Erlangen-N\"urnberg, Germany}
\date{\today}

\begin{abstract}
  We study the superconducting gap function of Sr$_2$RuO$_4$. 
  By solving the linearized Eliashberg equation with a correlated pairing vertex extracted from a dynamical mean-field calculation 
  we identify the dominant pairing channels. An analysis of the candidate gap functions in orbital and quasiparticle band basis reveals that an inter-orbital singlet pairing of even parity is in agreement with experimental observations. It reconciles in particular the occurrence of a two-component order parameter with the presence of line-nodes of quasiparticles along the c-axis in the superconducting phase. 
  The strong angular dependence of the gap along the Fermi surface is in stark contrast to its quasi-locality when expressed in the orbital basis. We identify local inter-orbital spin correlations as the driving force for the pairing and thus reveal the continuation of Hund's physics into the superconducting phase.
\end{abstract}
\maketitle


\section{Introduction}
Since its discovery 27 years ago \cite{Maeno1994}, 
the superconducting ground state of \stront remains a major challenge
in strongly correlated electron physics \cite{Bergemann2003, Mackenzie2003}. The symmetry of the order parameter is still a debated question, with recent experiments challenging established views~\cite{Pustogow2019b,Ishida2020} - for recent reviews and discussion see e.g. \cite{Mackenzie2020,A.P.MackenzieT.Scaffidi1965a,Kivelson2020}.

In the normal state, material-realistic calculations combining
density-functional theory (DFT) and dynamical mean-field theory (DMFT \cite{Georges1996,Kotliar2006}) 
have been able to explain, on a microscopic level, many experimental observations.
This includes the emergence of quasiparticles with anisotropic mass
renormalization when cooling down from the incoherent high temperature phase to
the Hund's metal Fermi liquid regime \cite{Mravlje2011} as well as quasiparticle dispersions and the subtle reshaping of the Fermi surface due to the spin-orbit-coupling as probed by photoemission experiments \cite{Tamai2019}.
More recently, momentum resolved static magnetic susceptibilities computed in
DMFT \cite{Boehnke2018, Strand2019a, Acharya2019} have been shown to agree well
with inelastic neutron-scattering experiments \cite{Steffens2019a}, when vertex
corrections are properly taken into account \cite{Boehnke2018, Strand2019a}.
These recent works highlight the correlated nature of \stront as a
Hund's metal~\cite{Georges_Hund_review_2013} and the importance of local spin-fluctuations, 
as in other Hund's metals like the iron-based 
superconductors \cite{Haule2009, Hansmann2010, Yin2011, Toschi2012}.

In contrast with the normal state, the nature of the low temperature superconducting state
of \stront remains a matter of intense debate today. Despite a large body of experiments,
no consensus has been reached for example on the precise form and symmetry of the
superconducting order.  The difficulty comes from the multiorbital nature of
this material and the spin-orbit-coupling (SOC), which yields a large number of
possible pairing symmetries (see Ref.~\onlinecite{Kaba2019b} for a classification based on group theory).
This makes a direct phenomenological approach difficult, in which one considers every
possible pairings authorized by symmetry and finds the (hopefully) unique one
compatible with every experiment. Several works have been published in this direction,
see e.g.  \cite{Ramires2016, Ramires2019b, Kivelson2020}.

In this work, we instead use microscopic computations at high temperature, in
the normal state Hund's metal regime, to guide us towards the most likely superconducting order
candidate. We directly solve the linearized Eliashberg equation for
superconductivity from high temperatures, non perturbatively, without any
assumption on the symmetry of the order parameter, for the same realistic model
used in the successful studies on the normal state. We identify two dominant
channels in the Eliashberg analysis. Both are {\it two-component spin-singlet}
modes. The first one has \emph{inter}-orbital character, even momentum- and
odd orbital symmetry, while the second has \emph{intra}-orbital character and
odd momentum symmetry. We will show, however, that the \emph{inter}-orbital channel
is the only one compatible with experiments. This kind of pairing was not identified in similar previous studies that used an RPA-like approximation \cite{Gingras2019} or a self-consistent GW + DMFT approach \cite{Acharya2019}.

This paper is organized as follows.  In Section \ref{sec:Model}, we present our
model, based on realistic electronic structure; in section \ref{sec:formalism}, we
present our formalism and discuss every approximation we make in this work at
a general level; in Sec. \ref{sec:results}, we discuss our numerical results
and the two dominant superconducting symmetry channels and compare to previous 
related works \cite{Gingras2019, Acharya2019}; in Sec. \ref{sec:experiments}, we present a detailed comparison of these two pairing states to experiments; finally we conclude in Sec.~\ref{sec:conclusion}.

\section{Model}
\label{sec:Model}
\begin{figure}[t]
  \includegraphics[width=\columnwidth]{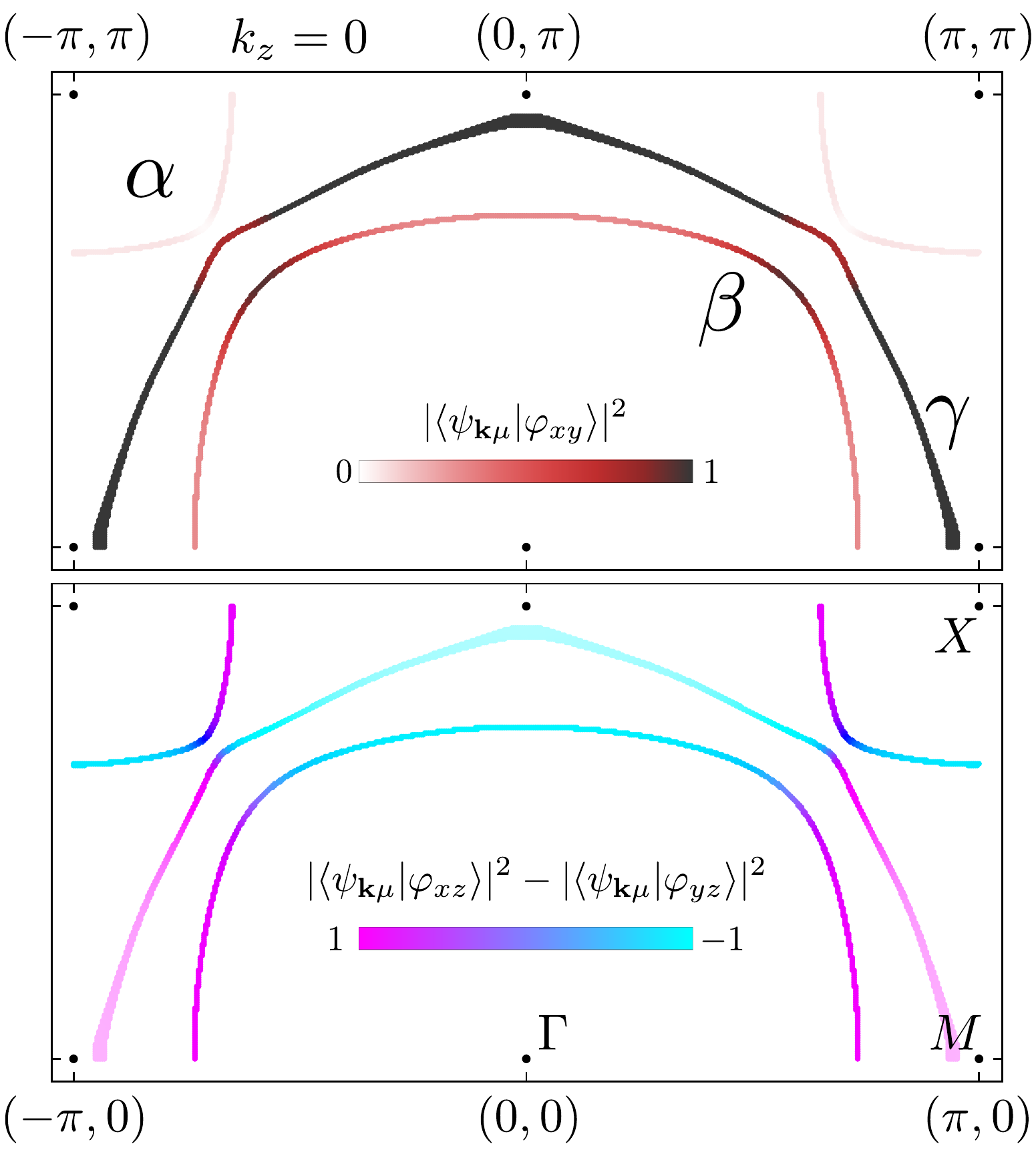}   
  \caption{Plot of the orbital character of the Fermi surface in the $k_x,k_y$ plane at $k_z=0$, 
  including spin-orbit coupling. The in-plane $xy$ orbital dominates the $\gamma$ Fermi surface sheet (top) while the out-of-plane orbitals $xz$ and $yz$ dominates the outer $\alpha$ hole-pocket and the inner $\beta$ electron pocket (bottom).
  }
  \label{fig:FSorb}
\end{figure}

The structure of Sr$_2$RuO$_4$ is that of an undistorted layered perovskite with $D4h$ tetragonal pointgroup symmetry on the Ru sites.
Due to the strong induced cubic crystal-field splitting of the Ru-4d states ($10Dq=E_{e_{\mathrm{g}}}-E_{t_{2\mathrm{g}}}$) the $e_{\mathrm{g}}$ states are pushed well above the Fermi level. Hence, the three bands crossing the Fermi level, can be represented by a localized Wannier-orbital basis transforming as the Ru-$4d$-$t_{2\mathrm{g}}$ orbitals $xy$, $xz$, and $yz$, with a nominal filling of four valence electrons.
While the tetragonal onsite splitting of these orbitals is rather small, $E_{xy}-E_{xz/yz}\approx 80\,$meV, the hopping parameters are remarkably anisotropic showing a 2D like dispersion for $xy$-states and 1D dispersion for $xz$ and $yz$-bands respectively. 
This can also be seen in the Fermi surface topology where the $xz$- and $yz$-states form nearly one dimensional Fermi surface sheets, that hybridize weakly and give rise to the hole-pocket $\alpha$ (centered around $X$) and the electron pocket $\beta$ (centered around $\Gamma$); For resolved orbital contributions of $xz$- and $yz$-states see the color coded lower panel of Fig. \ref{fig:FSorb}. The in-plane $xy$ states, on the other hand, is dominant in the formation of the electron pocket $\gamma$ (concentric with $\beta$), see Fig.~\ref{fig:FSorb} (upper panel).

Besides $D4h$ hopping and crystal-field potential, spin-orbit-coupling (SOC) effects are sizable for the Ru-$4d$ valence electrons and play an important role in \stront \cite{Haverkort2008,Veenstra2014}. For the three-band $t_{2\mathrm{g}}$ low-energy model it has been shown that the effect of SOC can be captured using a local $\mathbf{l}\cdot\mathbf{s}$ operator
and that the effective spin-orbit coupling is enhanced by correlations \cite{Zhang2016,Kim2018,Tamai2019}. Its effects within the $t_{2\mathrm{g}}$ subspace have significant impact in large parts of the Brillouin zone (BZ) around the Fermi level and, more specifically, the shape \emph{and orbital character} of the Fermi surface at the points where the Fermi-surface sheets $\alpha$, $\beta$, and $\gamma$ are in close proximity \cite{Tamai2019}, e.g.~along $\Gamma$--$X$ in Fig.~\ref{fig:FSorb}.

In this study the normal-state electronic structure of Sr$_2$RuO$_4$ is described using an ab initio derived low-energy $t_{2\mathrm{g}}$ effective model, computed using density-functional theory (DFT) and a Wannierizaton of the three Kohn-Sham bands crossing the Fermi level. The model has been used in several previous studies \cite{Mravlje2011, Tamai2019, Strand2019a}, for details on the model construction using Wien2k \cite{WIEN2k} and Wannier90 \cite{Kunes2010, *Marzari1997, *Mostofi2008, *Marzari2012} we refer to Ref.~\cite{Tamai2019}. The effective screened Coulomb interaction was modeled using the rotationally-invariant Kanamori form projected on the $t_{2\mathrm{g}}$ subspace, and parametrized by a Hubbard $U$ and a Hund's $J$. The interaction parameters $U=2.3\,\mathrm{eV}$ and $J=0.4\,\mathrm{eV}$ are set in accordance with previous studies, which have correctly reproduced quantum oscillation experiments \cite{Mravlje2011}, angle resolved photo emission spectroscopy \cite{Tamai2019}, 
and the momentum-dependent static spin response function \cite{Strand2019a}.\\

\section{Formalism} 
\label{sec:formalism}
In order to study the superconducting instability in Sr$_2$RuO$_4$, we solve the linearized Eliashberg equation in all possible
symmetry channels. We use a DMFT based approximation detailed below to approximate the
two body quantities involved in this equation. A more conventional approach to study superconductivity in (cluster-) DMFT would be
to directly solve the DMFT equations for a single-site or a small cluster in the ordered phased, 
or to compute the superconducting susceptibility
above $T_c$ from the linear response of {\it  one-body} quantities to a small pairing field.
This is well known in e.g. the study of $s-$wave or $d-$wave superconducting orders in attractive or repulsive Hubbard model,
see e.g. \cite{PhysRevLett.110.216405}.
The Eliashberg approach has the advantage of being able to describe an arbitrary momentum dependence of the superconducting gap
function $\Delta$, a very important question that can not be addressed with cluster DMFT methods.
However, this flexibility comes at a cost: a limitation to high temperatures and the need to rely on approximations for
two-body quantities which are only partially controled.

In this section, we first define the generalized propagators and self-energies in a Nambu basis as well as their symmetry
properties. Then we present the Eliashberg equation and the relevant Parquet and Bethe-Salpeter equations for the vertex.
Finally we discuss in detail our approximation strategy and the role of spin-orbit coupling for our results.

\subsection{Generalized propagators and self-energies}
First, we define the anomalous Green's functions at imaginary time $\tau$ and momentum $\mathbf{k}$ by 
\begin{align}
  \label{eq:SCorder1}
  F^{\sigma \sigma'}_{ab}(\mathbf{k}, \tau)\equiv \langle \mathcal{T}_\tau c^{}_{a\sigma}(\mathbf{k}, \tau)c^{}_{b\sigma'}(-\mathbf{k}, 0) \rangle
  \\
  \label{eq:SCorder2}
  \overline{F}^{\sigma \sigma'}_{ab}(\mathbf{k}, \tau)\equiv \langle \mathcal{T}_\tau c^{\dagger}_{a\sigma}(\mathbf{k}, \tau)c^{\dagger}_{b\sigma'}(-\mathbf{k}, 0) \rangle
\end{align}
where $\mathcal{T}_\tau$ is the time ordering operator and $c_{a\sigma}(\mathbf{k}, \tau)$ ($c^\dagger_{a\sigma}(\mathbf{k}, \tau)$)
are fermionic annihilation (creation) operators in the Heisenberg representation1 with orbital ($a$, $b$) and spin ($\sigma,\,\sigma' \in
\{\uparrow,\downarrow\}$) indices.
In a Nambu basis spanned by the spinors
\begin{multline}
  \mathbf{\Psi}_a(\mathbf{k}, \tau)
  \\ \equiv
  \begin{pmatrix}
    c^{}_{a\uparrow}(\mathbf{k}, \tau) \,,
    c^{}_{a\downarrow}(\mathbf{k}, \tau) \,,
    c^{\dagger}_{a\uparrow}(-\mathbf{k}, \tau) \,,
    c^{\dagger}_{a\downarrow}(-\mathbf{k}, \tau)
  \end{pmatrix}\, ,
  \label{eq:Nambu_spinor}
\end{multline}
the anomalous propagators are the block off-diagonal entries of the generalized single-particle propagator
\begin{align}
  \begin{split}
    \hat{\mathbf{G}}_{ab}(\mathbf{k}, \tau)
    &\equiv
    -\left\langle
      \mathcal{T}_\tau
      \mathbf{\Psi}_a(\mathbf{k}, \tau)
      \mathbf{\Psi}^\dagger_b(\mathbf{k}, 0)
    \right\rangle\\
    &=
    \begin{pmatrix}
      \mathbf{G}_{ab}(\mathbf{k}, \tau) & -\mathbf{F}_{ab}(\mathbf{k}, \tau) \\
      -\overline{\mathbf{F}}_{ab}(\mathbf{k}, \tau)& \overline{\mathbf{G}}_{ab}(\mathbf{k}, \tau)
    \end{pmatrix}\,.                             
  \end{split}
 \label{eq:general_G}
\end{align}
In this equation, $\mathbf{G}$ and $\mathbf{F}$ are $2\times2$ matrices in spin space.
The diagonal entries are the normal single-particle (single-hole) propagators
$\mathbf{G}_{ab}(\mathbf{k}, \tau)$ ($\overline{\mathbf{G}}_{ab}(\mathbf{k}, \tau)$). After
Fourier transforming from imaginary time to Matsubara frequency, the generalization of the
corresponding Dyson equation reads
\begin{equation}
  \hat{\mathbf{G}}(\mathbf{k}, i\nu_n) = \hat{\mathbf{G}}^{0}(\mathbf{k}, i\nu_n)+\hat{\mathbf{G}}^{0}(\mathbf{k}, i\nu_n)
	\mathbf{\Sigma}(\mathbf{k}, i\nu_n)
	\hat{\mathbf{G}}(\mathbf{k}, i\nu_n)
	\,,
	\label{eq:dyson_gorkov_general}
\end{equation}
where
\begin{align}
	\hat{\mathbf{G}}^{0}_{ab}(\mathbf{k}, i\nu_n)=
	\begin{pmatrix}
	\mathbf{G}^{0}_{ab}(\mathbf{k}, i\nu_n) & 0 \\
	0 & \overline{\mathbf{G}}^{0}_{ab}(\mathbf{k}, i\nu_n)
	\end{pmatrix}
            \,,
            \label{eq:G0}
\end{align}
is the non-interacting generalized propagator and
\begin{align}
  \mathbf{\Sigma}_{ab}(\mathbf{k}, i\nu_n)=
  \begin{pmatrix}
    \mathbf{\Sigma}^\text{norm.}_{ab}(\mathbf{k}, i\nu_n) & \mathbf{\Delta}_{ab}(\mathbf{k}, i\nu_n) \\
    \overline{\mathbf{\Delta}}_{ab}(\mathbf{k}, i\nu_n) & \overline{\mathbf{\Sigma}}^\text{norm.}_{ab}(\mathbf{k}, i\nu_n)
  \end{pmatrix}\,,
  \label{eq:sigma_general}
\end{align}
defines the generalized self-energy. As for the Green's function $\hat{\mathbf{G}}$, the diagonal blocks in the self-energy, $\mathbf{\Sigma}^{\text{norm.}}$ and $\overline{\mathbf{\Sigma}}^{\text{norm.}}$, are the single-particle and single-hole self energies and the off-diagonal pairing term $\Delta$ is called \emph{the super-conducting gap function}. While $\Delta$ in principle only is finite in the symmetry broken super-conducting state, it is possible to determine the super-conducting instability in terms of $\Delta$ directly from the normal-phase using the linearized Eliashberg formalism.

\subsection{The Eliashberg and Parquet equations}
\label{subsec:eliashberg}
The linearized Eliashberg equation for the superconducting gap function $\Delta$ reads 
\cite{Eliashberg1960, Abrikosov, Nourafkan2016a, Gingras2019}
\begin{multline}
    \lambda \Delta^{\text{s/t}}_{ab}(K) =
    -\frac{1}{2N_\mathbf{k}\beta} \sum_{K'} \Gamma^{\text{s/t}}_{cadb}(Q=0, K', K)\times\\
     G_{cf}(-K') G_{de}(K') \Delta^{\text{s/t}}_{ef}(K')\,,
  \label{eq:eliashberg}
\end{multline}
where $\Gamma^{\text{s/t}}$ is the irreducible vertex function in the singlet/triplet channel, $G$ is the spin-independent (interacting) single-particle lattice Green function, $Q/K$ is the bosonic/fermionic four-vector
consisting of Matsubara frequency ($\omega_n/\nu_n$) and momentum ($\mathbf{q}/\mathbf{k}$), $N_{\mathbf{k}}$ is
the number of momenta and $\beta$ is the inverse temperature. Here all Latin indices are orbital indices and
the Einstein summation convention is used.

The Eliashberg equation [Eq.~\eqref{eq:eliashberg}] is an eigenvalue problem with eigenvectors $\Delta$ and eigenvalues $\lambda$.
When lowering the temperature, the superconducting instability
occurs at the critical temperature $T_c$, at which the eigenvalue $\lambda$ of the dominant eigenvector gap function $\Delta$ reaches unity.
In practice, because our numerical solution is limited to high and intermediate
temperatures, we are not able to reach $T_c$, but we will rather compare the evolution of the ten largest eigenvalues $\lambda$ and the corresponding eigenvector gap functions $\Delta$ as a function of temperature.

In this paper, we neglect the spin-orbit contributions to the pairing vertex, which allows us to solve Eq.~\eqref{eq:eliashberg} separately for singlet (s) and triplet (t) channels in the spin diagonalized form, i.e.
\begin{align}
	\Delta^{\text{s/t}}_{ab}
	=
	\Delta_{ab}^{\uparrow\downarrow}
	\mp
	\Delta_{ab}^{\downarrow\uparrow}
	\,,
	\quad
	\Gamma^{\text{s/t}}_{abcd}
    =
    \Gamma_{abcd}^{\uparrow\uparrow\downarrow\downarrow}
    \mp
    \Gamma_{abcd}^{\uparrow\downarrow\downarrow\uparrow}
	\,.
\end{align}
The irreducible vertex $\Gamma^{\text{s/t}}$  can be computed from the fully irreducible
vertex $\Lambda$ and the reducible vertex $\Phi^{\mathrm{d/m}}$ through the Parquet equation \cite{RevModPhys.90.025003, Bickers2006, Rohringer}.
The irreducible singlet vertex $\Gamma^{\text{s}}$ is given by
\begin{align}
\begin{split}
	\Gamma^{\text{s}}_{abcd}(Q, K, K') =
	-
	\Lambda^{\text{s}}_{abcd}(Q, K, K')
	\\
	+
	\left[
	\frac{3}{2}
	\Phi^{\text{m}}_{abcd}
	-
	\frac{1}{2}
	\Phi^{\text{d}}_{abcd}
	\right](Q-K-K', K, K')\\
	+
	\left[
	\frac{3}{2}
	\Phi^{\text{m}}_{cbad}
	-
	\frac{1}{2}
	\Phi^{\text{d}}_{cbad}
	\right](K-K', Q-K, K')
	\,,
	\label{eq:gamma_singlet_no_approximation}
\end{split}
\end{align}
and the irreducible triplet vertex $\Gamma^{\text{t}}$ by
\begin{align}
\begin{split}
\Gamma^{\text{t}}_{abcd}(Q, K, K') =
\Lambda^{\text{t}}_{abcd}(Q, K, K')
\\
+
\left[
\frac{1}{2}
\Phi^{\text{m}}_{abcd}
+
\frac{1}{2}
\Phi^{\text{d}}_{abcd}
\right](Q-K-K', K, K')\\
-
\left[
\frac{1}{2}
\Phi^{\text{m}}_{cbad}
+
\frac{1}{2}
\Phi^{\text{d}}_{cbad}
\right](K-K', Q-K, K')
\,,
	\label{eq:gamma_triplet_no_approximation}
\end{split}
\end{align}
where the superscript $\mathrm{d/m}$ indicates the density/magnetic channel \cite{RevModPhys.90.025003},
$\Phi^{\text{d/m}}_{abcd} =
\Phi_{abcd}^{\uparrow\uparrow\uparrow\uparrow} \pm	\Phi_{abcd}^{\uparrow\uparrow\downarrow\downarrow}$.

The reducible vertex function $\Phi^{\mathrm{d/m}}$ is related to $\Gamma^{\mathrm{d/m}}$ the irreducible vertex function in the density/magnetic channel by 
\begin{multline}
\label{eq:laddervertex}
\Phi^{\text{d/m}}_{abcd}(Q, K, K')
=
\frac{1}{(N_\mathbf{k}\beta)^2}
\sum_{P_1, P_2}
\Gamma^{\text{d/m}}_{abef}(Q, K, P_1)
\times
\\
\chi^{\text{d/m}}_{fegh}(Q, P_1,P_2)
\Gamma^{\text{d/m}}_{hgcd}(Q, P_2, K')
\, ,
\end{multline}
Here, $\chi^{\mathrm{d/m}}$ is the generalized susceptibility, obtained by solving the Bethe-Salpeter equation
   \begin{multline}
   \label{eq:bse_approx}
      \chi_{abcd}^{\text{d/m}}(Q, K, K') 
      =\chi_{abcd}^{0,\mathrm{d/m}}(Q, K, K')
     + \\
      \frac{1}{(N_{\mathbf{k}}\beta)^2}
      \sum_{P_1 P_2}
      \chi_{abef}^{0,\mathrm{d/m}}(Q, K, P_1)
      \Gamma_{fegh}^{\text{d/m}}(Q, P_1, P_2)
      \times
      \\
      \chi_{hgcd}^{\text{d/m}}(Q, P_2, K')
   \end{multline}
where $\chi^{0,\mathrm{d/m}}$ is the bare lattice ``bubble'':
\begin{equation}
\chi_{abcd}^{0,\mathrm{d/m}}(Q, K, K')
\equiv
-N_{\mathbf{k}} \beta G_{da}(K)G_{bc}(K+Q) \delta_{K, K'}
\,.
\end{equation}

\subsection{Approximation strategy}

The Eliashberg and Parquet equations written above are exact, but
untractable.  Hence we develop an approximation based on the
DMFT solution of this material.

Let us first emphasize the central role played in these equations by the
magnetic susceptibility $\chi^{\mathrm{m}}$, as we see in Eq. \eqref{eq:laddervertex}. 
In Ref.~\onlinecite{Strand2019a}, $\chi^{\mathrm{m}}$ has been studied for this material in
great details within the DMFT approximation, {\it including vertex corrections}. 
An excellent agreement with experiments was obtained, but only if vertex corrections were included through the Bethe-Salpeter equation [Eq.~\eqref{eq:bse_approx}]. 
Our approximation strategy consist therefore in using the full DMFT generalized susceptibility $\chi^{\mathrm{d/m}}_{abcd}$ in Eq.~\eqref{eq:laddervertex} (with vertex corrections) and approximate the other
vertices $\Lambda$ and $\Gamma$ with simple static (effective) approximations.

More precisely, we first approximate the fully irreducible vertex $\Lambda$ by its bare value $\tilde{\Lambda}$
\begin{equation}
   \Lambda \approx \tilde{\Lambda}
\end{equation}
whose exact form is given in App.~\ref{app:ApproxFullIrrVertex} as a function of $U$ and $J$.

Secondly, we approximate $\Gamma^\text{d/m}$ by a static effective interaction.
Conceptually, we could use here the DMFT impurity vertex
$\Gamma^\text{d/m}_\text{DMFT}$, but as explained in details in
App.~\ref{app:StaticApproxVertexLadder}, we replace it by an effective
static approximation which reproduces well the low energy behaviour of
$\Gamma^{s}$. This is necessary in practice because of the stochastic noise in
$\Lambda$ and $\Gamma$ in the hybridization expansion quantum Monte Carlo
(CT-HYB) \cite{Werner2006, Werner2006a, Haule2007, Gull2011} solver we use here
to solve the DMFT self-consistent impurity model.

\subsection{The role of spin-orbit-coupling}
\label{subsec:SOC} 
As has been pointed out in previous works (e.g. \cite{Haverkort2008,Veenstra2014}) and as mentioned in Sec.~\ref{sec:Model}, spin-orbit-coupling affects the shape and orbital character of the Fermi surface significantly. In its presence, spin ceases to be a good quantum number for the quasiparticles around $\varepsilon_F$, the anomalous propagators of the superconducting phase Eqs. \eqref{eq:SCorder1} and \eqref{eq:SCorder2} are not diagonal in the spin eigenbasis, and Cooper pairs cannot be classified as either singlet- or triplet. Moreover, the strong momentum dependence of the SOC entanglement prohibits any transform to a local pseudospin for recovery of singlet- and triplet notion (different, e.g., from the local $\mathbf{J}^2,J_z$ basis in Ce based heavy-fermion superconductors).

While there have been weak coupling studies that include spin-orbit coupling \cite{Zhang2018i, Romer2019b, Scaffidi2014} on the level of the pairing vertex, including SOC in our non-perturbative approach on the level of the Parquet equations Eqs. \eqref{eq:gamma_singlet_no_approximation} and \eqref{eq:gamma_triplet_no_approximation} is currently not feasible in practice. We therefore employ an approximation strategy, which has previously been successfully applied for the calculation of the magnetic response \cite{Strand2019a} and neglect SOC on the level of the two particle vertices. This means, that our anomalous self energy $\Delta$ (Eq.~\eqref{eq:sigma_general}) \emph{can be classified as singlet or triplet} in nature. For the anomalous propagator $\mathbf{F}$, however, the SOC entanglement enters via the normal state propagator $\mathbf{G}^0$ Eq.~\eqref{eq:G0}. 

\section{Results}
\label{sec:results}
\subsection{Solutions of the Eliashberg equation}
\begin{figure}[!htb]
  \includegraphics{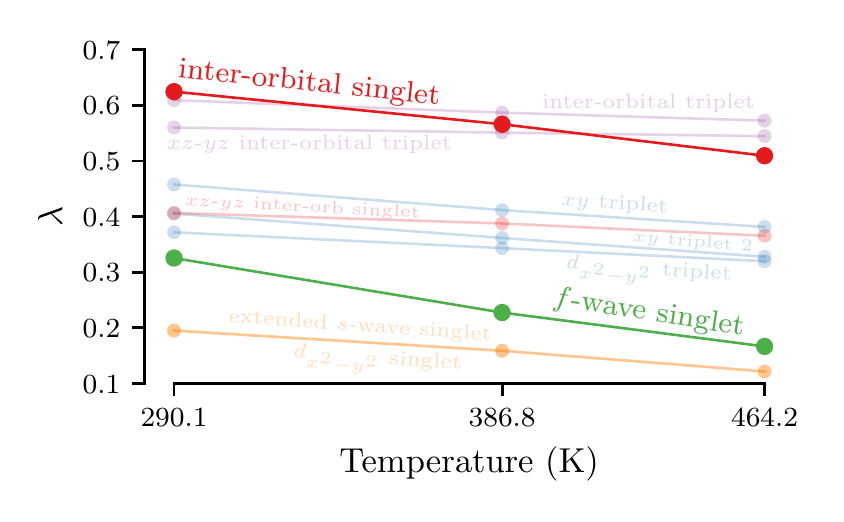}   
  \caption{Plot of the temperature dependent Eliashberg $\lambda$ eigenvalues for the ten gap functions closest to $\lambda=1$. We highlight the two selected prime candidates - an inter-orbital singlet (red) and an $f$-wave singlet (green) - for our further analysis.}
  \label{fig:dmft_lambda_over_temperature}
\end{figure}
\begin{table*}[!htb]
\begin{threeparttable}
	\renewcommand{\arraystretch}{1.7}
	\caption{The ten leading gap functions in descending order at $T\approx290\,\text{K}$ (c.f. Fig. \ref{fig:dmft_lambda_over_temperature}). Our considered prime candidates $\Delta^{i.o.}$ and $\Delta^{f}$ are highlighted in gray. Gaps with the same $SPOT$ symmetry share the same color. As many of our gap functions have not only a single finite matrix element in the orbital space the column ``Orbital character'' indicates the dominant orbital matrix element (see also Fig.~\ref{fig:gap_candidates_in_different_bases}).}
	\begin{tabularx}{0.85\textwidth}{ >{\hsize=0.8\hsize}L<{\kern\tabcolsep}@{}
			>{\hsize=0.2\hsize}C<{\kern\tabcolsep}@{}
			>{\hsize=0.2\hsize}C<{\kern\tabcolsep}@{}
			>{\hsize=0.2\hsize}C<{\kern\tabcolsep}@{}
			>{\hsize=0.2\hsize}C<{\kern\tabcolsep}@{}
			>{\hsize=0.8\hsize}L<{\kern\tabcolsep}@{}}
		\hline\hline
		&\multicolumn{4}{c}{Symmetries} &\\[-.45cm]
		\cline{2-5} \\[-1cm]
		Pairing & Spin & Parity & Orbital & Time & \hspace{1.05cm} Orbital character  \\
		\hline
		\rowcolor[gray]{0.95}
		\textbf{inter-orbital singlet} $\Delta^{i.o.}$ & \textcolor{inter_orb_singlet_xyxz_xyyz}{$\bm{-}$} & \textcolor{inter_orb_singlet_xyxz_xyyz}{$\bm{+}$} & \textcolor{inter_orb_singlet_xyxz_xyyz}{$\bm{-}$} & \textcolor{inter_orb_singlet_xyxz_xyyz}{$\bm{-}$} &\hspace{1.06cm}degenerate $\begin{cases} \text{inter }xy\text{-}yz \\ \text{inter }xy\text{-}xz \end{cases}$ \\
		inter-orbital triplet & \textcolor{inter_orb_triplet_xyxz_xyyz}{$\bm{+}$} & \textcolor{inter_orb_triplet_xyxz_xyyz}{$\bm{+}$} & \textcolor{inter_orb_triplet_xyxz_xyyz}{$\bm{-}$} & \textcolor{inter_orb_triplet_xyxz_xyyz}{$\bm{+}$} &\hspace{1.06cm}degenerate $\begin{cases} \text{inter }xy\text{-}yz \\ \text{inter }xy\text{-}xz \end{cases}$ \\
		$xz$-$yz$ inter-orbital triplet & \textcolor{inter_orb_triplet_xyxz_xyyz}{$\bm{+}$} & \textcolor{inter_orb_triplet_xyxz_xyyz}{$\bm{+}$} & \textcolor{inter_orb_triplet_xyxz_xyyz}{$\bm{-}$} & \textcolor{inter_orb_triplet_xyxz_xyyz}{$\bm{+}$} &\hspace{2.85cm} inter $xz$-$yz$ \\
		$xy$ triplet & \textcolor{triplet_d_x2_y2}{$\bm{+}$} & \textcolor{triplet_d_x2_y2}{$\bm{+}$} & \textcolor{triplet_d_x2_y2}{$\bm{+}$} & \textcolor{triplet_d_x2_y2}{$\bm{-}$} &\hspace{2.85cm} intra $xy$ \\
		$xz$-$yz$ inter-orbital singlet & \textcolor{inter_orb_singlet_xyxz_xyyz}{$\bm{-}$} & \textcolor{inter_orb_singlet_xyxz_xyyz}{$\bm{+}$} & \textcolor{inter_orb_singlet_xyxz_xyyz}{$\bm{-}$} & \textcolor{inter_orb_singlet_xyxz_xyyz}{$\bm{-}$} & \hspace{2.85cm}  inter $xz$-$yz$ \\
		$xy$ triplet 2 &\textcolor{triplet_d_x2_y2}{$\bm{+}$} & \textcolor{triplet_d_x2_y2}{$\bm{+}$} & \textcolor{triplet_d_x2_y2}{$\bm{+}$} & \textcolor{triplet_d_x2_y2}{$\bm{-}$} &\hspace{2.85cm} intra $xy$ \\
		$d_{x^2-y^2}$ triplet & \textcolor{triplet_d_x2_y2}{$\bm{+}$} & \textcolor{triplet_d_x2_y2}{$\bm{+}$} & \textcolor{triplet_d_x2_y2}{$\bm{+}$} & \textcolor{triplet_d_x2_y2}{$\bm{-}$} &\hspace{2.85cm} intra $xz/yz$  \\
		\rowcolor[gray]{0.95}$\mathbf{f}$-\textbf{wave singlet} $\Delta^f$& \textcolor{singlet_f}{$\bm{-}$} & \textcolor{singlet_f}{$\bm{-}$} & \textcolor{singlet_f}{$\bm{+}$} & \textcolor{singlet_f}{$\bm{-}$} &\hspace{1cm} degenerate $\begin{cases} \text{intra } xz \\ \text{intra }yz \end{cases}$ \\
		extended $s$-wave singlet & \textcolor{singlet_d_x2_y2}{$\bm{-}$} & \textcolor{singlet_d_x2_y2}{$\bm{+}$} & \textcolor{singlet_d_x2_y2}{$\bm{+}$} & \textcolor{singlet_d_x2_y2}{$\bm{+}$} & \hspace{2.85cm}  intra $xz/yz$
		\\
		$d_{x^2-y^2}$ singlet & \textcolor{singlet_d_x2_y2}{$\bm{-}$} & \textcolor{singlet_d_x2_y2}{$\bm{+}$} & \textcolor{singlet_d_x2_y2}{$\bm{+}$} & \textcolor{singlet_d_x2_y2}{$\bm{+}$} & \hspace{2.85cm}  intra $xz/yz$
		\\
		\hline
		\hline
	\end{tabularx}
	\label{tab:dmft_leading_gaps}
\end{threeparttable}
\end{table*}
\begin{figure*}[!htbp]
  \subfloat[The two degenerate inter-orbital singlets $\Delta^{i.o.}_{xy-xz}$, $\Delta^{i.o.}_{xy-yz}$  in the orbital basis]{\includegraphics[scale=1]{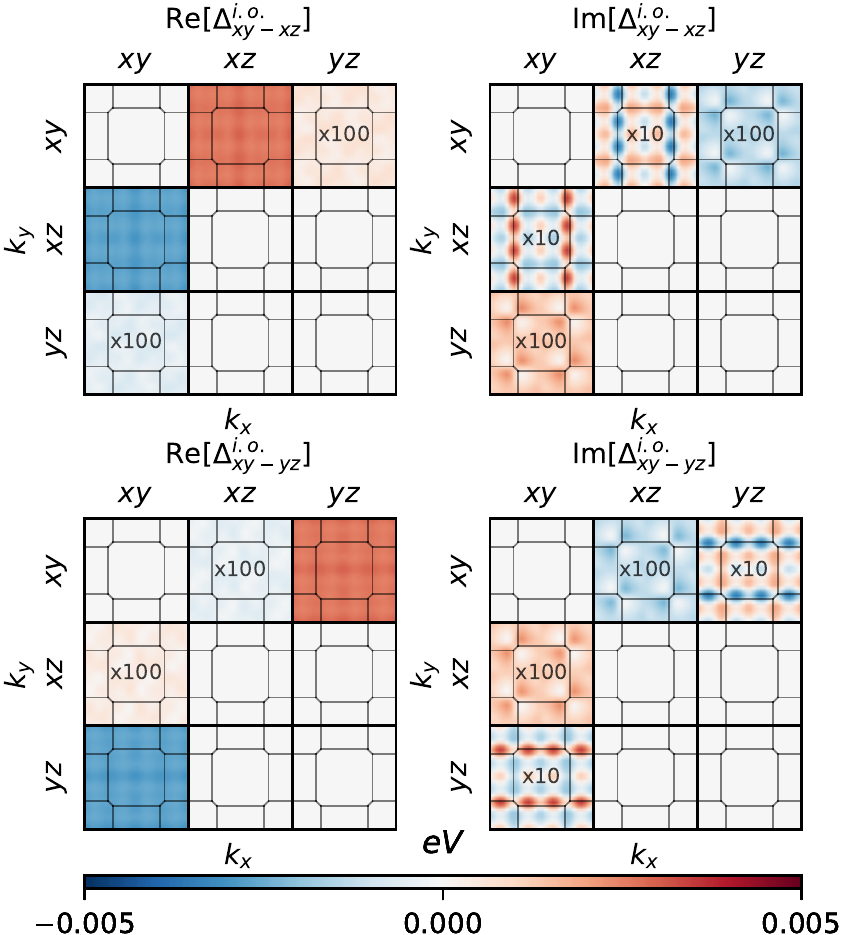}\label{fig:gap_candidates_in_different_bases_a} \ }
        \subfloat[The two degenerate $f$-wave singlets $\Delta^{f}_{xz}$, $\Delta^{f}_{yz}$ in the orbital basis]{\includegraphics[scale=1]{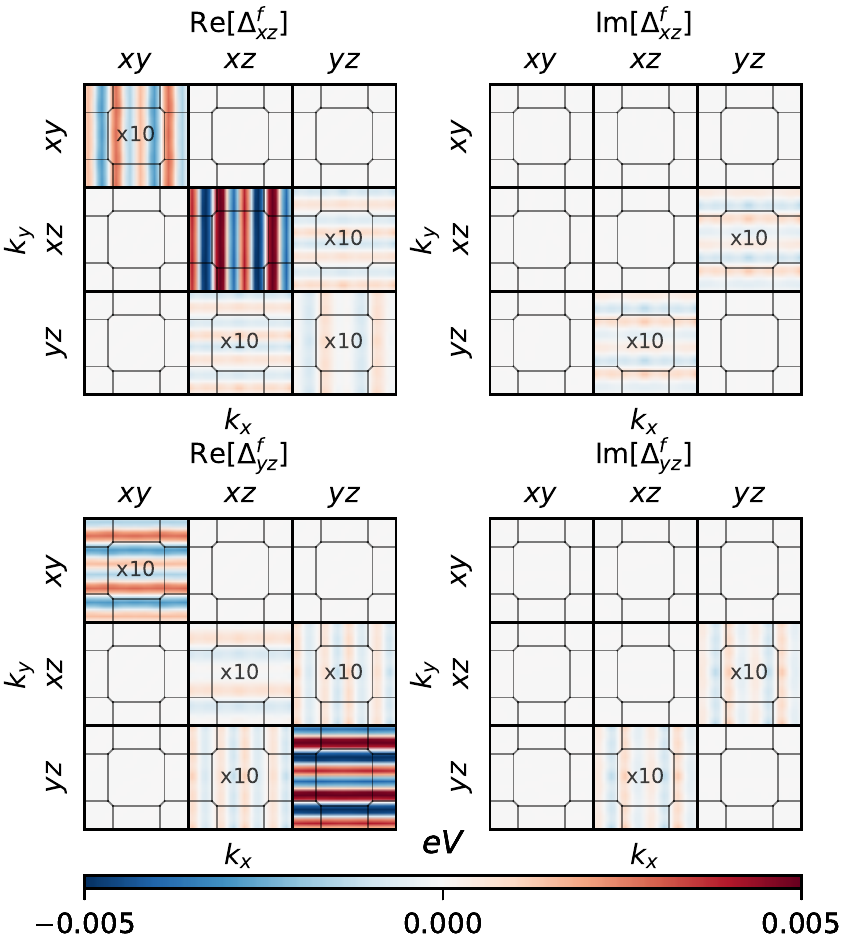}\label{fig:gap_candidates_in_different_bases_b} \ }
  \caption{Momentum dependence of the gap functions $\Delta_{ab}(k_x, k_y, k_z=0,i\nu_0)$ 
  in the $t_{2\mathrm{g}}$ orbital basis, $a,b \in \{xz, yz, xy\}$ at the first Matsubara frequency $i\nu_0$. Each entry of the matrices display the momentum-dependence as a function of $(k_x,k_y)$ in the Brillouin zone. The real and imaginary components are shown separately for, (a) the degenerate inter-orbital singlet gaps $\Delta^{i.o.}_{xy-xz}$ and $\Delta^{i.o.}_{xy-yz}$ with only orbital off-diagonal contributions, and (b) the degenerate intra-orbital $f$-wave singlet gaps $\Delta^{i.o.}_{zx}$ and $\Delta^{i.o.}_{yz}$. (Small orbital components are scaled with the scaling factors shown in the respective orbital component panel). 
}
  \label{fig:gap_candidates_in_different_bases}
\end{figure*}
\begin{figure*}[!htbp]
	\subfloat[The $xy-xz$ component of the inter-orbital singlet gap function.]{\includegraphics{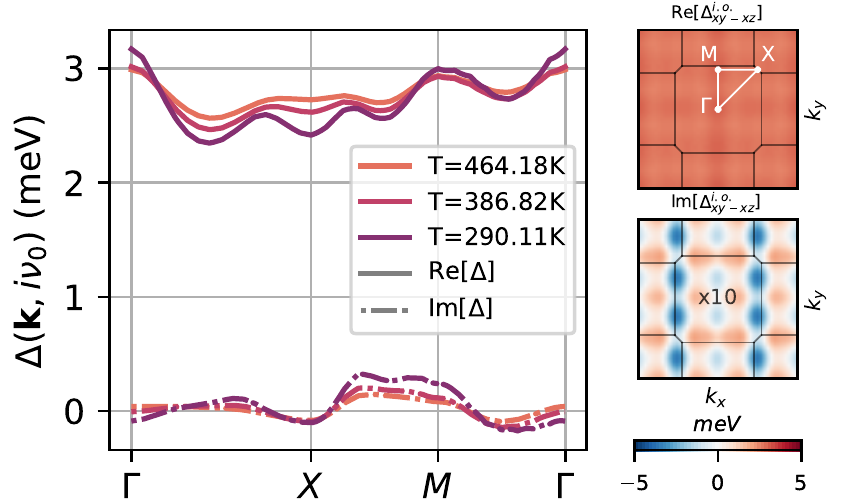} \label{fig:bandpath_interorb} \ }
	\subfloat[The $xz-xz$ component of the $f$-wave singlet gap function.]{\includegraphics{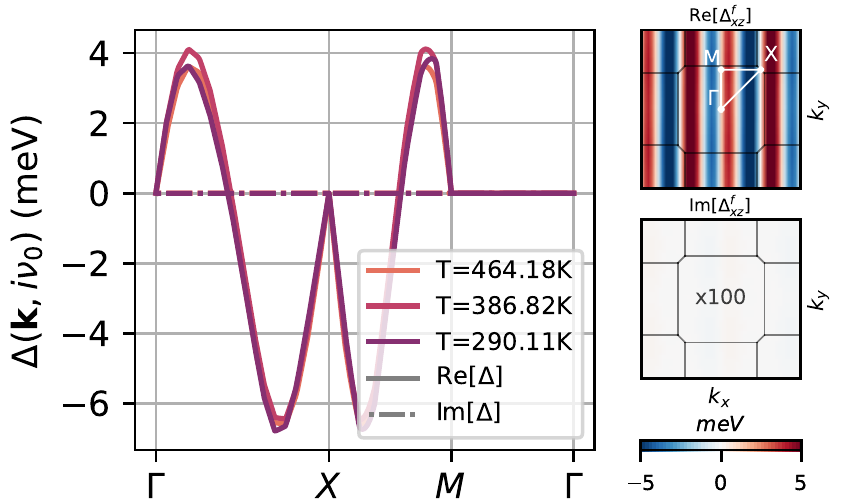} \label{fig:bandpath_f} \ }
	\caption{High symmetry k-space paths for the dominating orbital components in the two candidate gap functions for three temperatures. The k-space plane cuts are for the intermediate temperature and equivalent to Fig.~\ref{fig:gap_candidates_in_different_bases}.}
	\label{fig:bandpath_delta}
\end{figure*}

A finite superconducting gap is indicated by an eigenvalue $\lambda_i\geq 1$ in the linearized Eliashberg equation Eq.~\eqref{eq:eliashberg}. For the considered temperatures, all solutions of our scheme yield $\lambda_i<1$ which is expected. Nonetheless, the $\lambda_i$ eigenvalues above $T_c$ serve as indicators for potentially dominating gap symmetries at lower temperatures and at $T_c$. Hence, we sort the gap-functions in descending order for the calculated temperatures. We further classify them by their (anti-)symmetry upon permuting  spin, parity, orbital, and time indices.
\begin{subequations}
\begin{align}
	\hat{S}\Delta_{ab}^{\sigma\sigma'}(i\nu, \mathbf{k}) 
	&=
	 \Delta_{ab}^{\sigma'\sigma}(i\nu, \mathbf{k})\,,\\
	\hat{P}\Delta_{ab}^{\sigma\sigma'}(i\nu, \mathbf{k}) 
	&=
	 \Delta_{ab}^{\sigma\sigma'}(i\nu, -\mathbf{k})\,,\\
	\hat{O}\Delta_{ab}^{\sigma\sigma'}(i\nu, \mathbf{k}) 
	&=
	 \Delta_{ba}^{\sigma\sigma'}(i\nu, \mathbf{k})\,,\\
	\hat{T}\Delta_{ab}^{\sigma\sigma'}(i\nu, \mathbf{k}) 
	&=
	 \Delta_{ab}^{\sigma\sigma'}(-i\nu, \mathbf{k})\,.
\end{align}
\end{subequations}
where $a$, $b$ are orbital- and $\sigma$,$\sigma'$ are spin indices. The overall antisymmetry of the gap function which is dictated by the Pauli principle is usually formalized as the ``$SPOT$'' condition \cite{Linder2019, Gingras2019, Berezinskii1974a, Balatsky1992},
\begin{align}
  \hat{S}\hat{P}\hat{O}\hat{T} \Delta_{ab}^{\sigma\sigma'}(i\nu, \mathbf{k}) 
  =
  - \Delta_{ab}^{\sigma\sigma'}(i\nu, \mathbf{k})\,
  \label{eq:SPOT_def}
\end{align}
and - by construction - fulfilled by our gap functions. In Tab.~\ref{tab:dmft_leading_gaps} (which contains the gaps resolved $SPOT$ symmetry) we summarize our results for the ten gap functions with the highest $\lambda$ eigenvalues\footnote{This choice is motivated by the fact that the $d_{x^2-y^2}$ singlet gap function, which is leading in weak-coupling theory (see App.~\ref{sec:results_rpa}), dropped down to position ten in our non-perturbative scheme}. 

In Fig.~\ref{fig:dmft_lambda_over_temperature} the temperature dependence is highlighted which motivates our identification of prime candidates for the gap function in Sr$_2$RuO$_4$. At the lowest calculated temperatures, we find the largest $\lambda$ eigenvalue for a doubly degenerate inter-orbital ($xy$-$xz$,$xy$-$yz$) singlet. As it is the highest singlet with the second highest temperature gradient (red points in In Fig.~\ref{fig:dmft_lambda_over_temperature}) we will consider this inter-orbital gap function (from now on referred to as $\Delta^{i.o.}$) as our first principal candidate. As a second candidate we identify a doubly degenerate intra-orbital ($xz$ and $yz$) $f$-wave singlet ($\Delta^f$) which shows the overall strongest temperature gradient (green line in Fig.~\ref{fig:dmft_lambda_over_temperature}).
In Fig.~\ref{fig:gap_candidates_in_different_bases} we plot both components of $\Delta^{i.o.}$ (a) and $\Delta^f$ (b) as a matrix of their orbital indices. Each matrix element of the plot shows - as a color map - the momentum dependence of the corresponding gap function w.r.t. $k_x$ and $k_y$ in the $k_z=0$-plane. From these plots we understand directly the two-fold degeneracy of both gap functions as their components are transformed into one another upon the tetragonal symmetry $\pi/2$-rotation around the $k_z$-axis. The orbital and momentum dependence of $\Delta^{i.o.}$ and $\Delta^{f}$ is, however, completely different.

$\Delta^{i.o.}$ is odd under permutation of orbital indices but has even parity. Most remarkable, however, is its very weak $\mathbf{k}$-dependence of the dominant matrix elements (i.e. $\text{Re}[\Delta^{i.o.}_{xy-xz}]$ and  $\text{Re}[\Delta^{i.o.}_{xy-yz}]$) which displays a variation of less than $10\%$ around its mean value in the BZ. To quantify this we plot in Fig.~\ref{fig:bandpath_delta} (a) the $\mathbf{k}$-dependence of the dominant matrix elements of real and imaginary part of the gap function along a high symmetry path in the BZ for the three temperatures of our calculation. (Due to symmetry relations, we only need one plot for real and imaginary part each). The remarkably weak $\mathbf{k}$-dependence of $\Delta^{i.o.}$ in the orbital basis suggests a rather local pairing mechanism. If there is not a dramatic increase of $\mathbf{k}$-dependence at lower $T$ this is a promising outlook for DMFT studies. Computing the superconducting susceptibility by application of an inter-orbital pairing field would allow for an extrapolation to the lowest temperatures and eventually an estimate of $T_c$. 

We now turn to $\Delta^f$ which is orbitally practically diagonal. It is even w.r.t. orbital index permutations but - as the name suggests - odd in parity. It shows several horizontal nodal lines in the BZ. Its strong $\mathbf{k}$-dependence can be seen in detail in Fig.~\ref{fig:bandpath_delta} (b) and suggests a much more non-local pairing in real-space. Because of this strong momentum dependence, the direct study of this pairing state in the superconducting state with cluster-DMFT methods would require impractically large cluster sizes.
We stress here that the plots in Fig.~\ref{fig:gap_candidates_in_different_bases} are shown in the \emph{orbital basis} and must not be superimposed with the Fermi-surfaces which are strongly mixed in their orbital character (see Fig.~\ref{fig:FSorb}).

As there have been previous theoretical studies of pairing in \stront using the Eliashberg approach, we need to put our results into context. Some of the gaps listed as solutions in Tab.~\ref{tab:dmft_leading_gaps} have been discussed as potential candidates for Sr$_2$RuO$_4$ in the context of an RPA-like approximation \cite{Gingras2019} (which we reproduce in App. \ref{sec:results_rpa}). The overall discrepancy between our results and such a weak-coupling approach is expected however, as vertex corrections strongly affect the pairing mechanism (see next section). Indeed, it has been shown~\cite{Strand2019a} that including the full vertex when calculating the momentum-dependent magnetic response 
function yields a much better agreement with neutron scattering experiments~\cite{Steffens2019a} and corrects some of the significant discrepancies found at the RPA level. The closest study to our approach is a self-consistent GW + DMFT \cite{Acharya2019} study with which we have, despite differences in the approximations, an overall good agreement. The dominant pairing states that we find in the present work were however not identified or addressed in Ref.~\cite{Acharya2019}, which we attribute to the symmetry constraints imposed in that study.\\

We stress that our results are obtained above $T_c$ and should be understood as a \emph{proposal} for possible gap-symmetry and pairing mechanism in Sr$_2$RuO$_4$ and not as an ``\emph{ab initio} prediction''. Indeed we cannot judge with certainty how the eigenvalues will change at lower temperatures and if, e.g., the $f$-wave singlet will surpass the inter-orbital singlet. As we will show later, however, comparison to the large amount of experimental evidence lets us conclude that the inter-orbital singlet gap can be i) reconciled with all experiments and ii) could even help to explain an unsolved puzzle in quasiparticle interference (QPI) experiments\cite{Sharma2020}.

\begin{figure*}[!htbp]
	\includegraphics{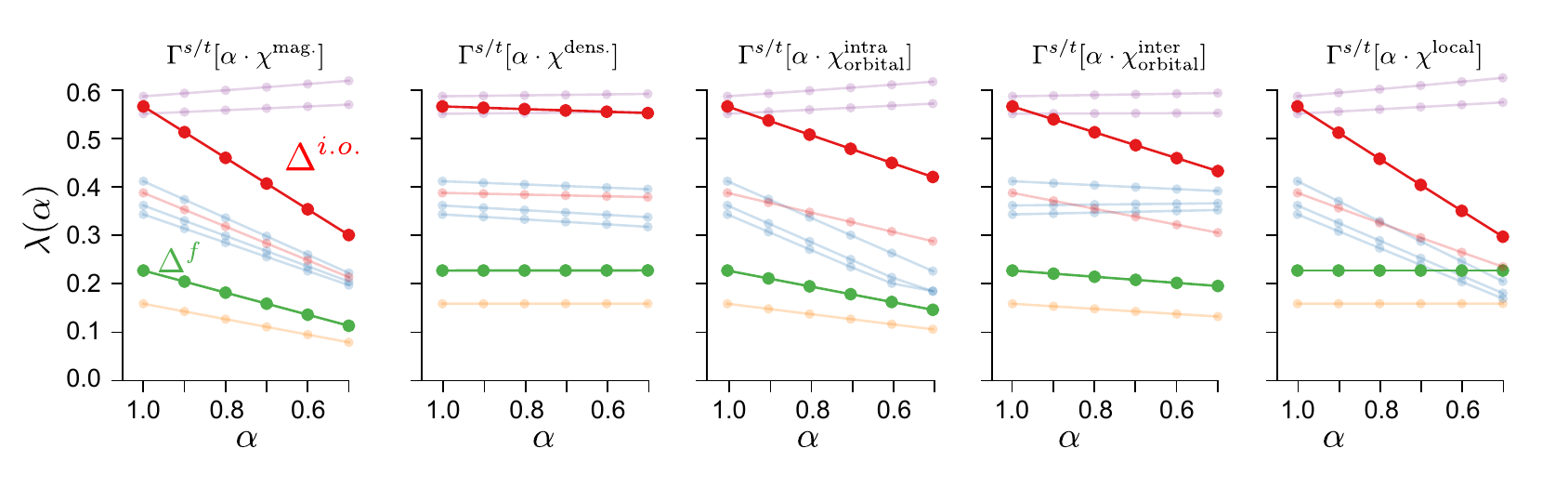}
	\caption{Dependence of the Eliashberg $\lambda$ eigenvalues at $T\approx386\,\mathrm{K}$ on the scaling factor $\alpha$ tuning different fluctuation channels in the pairing vertex.}
	\label{fig:channel_analysis}
\end{figure*}

\subsection{Channel analysis}
\label{channels}
In order to shed further light on the pairing mechanism of the gap functions reported in Tab.~\ref{tab:dmft_leading_gaps} we perform a ``channel analysis'' to determine which two-particle correlators drive the pairing for each of the gap functions. To do this, we down-scale selected components of the lattice susceptibility which enters the equation for the pairing vertices Eqs. \eqref{eq:gamma_singlet_no_approximation} and \eqref{eq:gamma_triplet_no_approximation}. Specifically, we distinguish between magnetic-, density-, inter-/intra-orbital-, and local fluctuations. For details see App. \ref{appendix:scaling_chi}. Keeping the original gap functions fixed, we compute new $\lambda$-values with the modified pairing vertices $\Gamma^{\mathrm{s/t}}$. The results for $T\approx386\,\mathrm{K}$ are shown in Fig.~\ref{fig:channel_analysis}. Concentrating on our prime candidates $\Delta^{i.o.}$ and $\Delta^{f}$, we note that both are mainly driven by spin-fluctuations. Especially $\Delta^{i.o.}$ seems to be strongly driven by magnetic fluctuations and, more specifically, by local magnetic inter- and intra-orbital correlations. Given that these correlations dominate also the Hund's metallic normal state, it seems plausible that a superconducting gap function driven by such correlations turns out to have a large Eliashberg eigenvalue. We note that superconductivity due to local spin fluctuations in the context of Hund's metals has been studied in a different context, e.g. in Ref.~\cite{Hoshino2015} in relation to spin-triplet pairing and in Ref.~\cite{Lee2018} in relation to the non-Fermi liquid spin dynamics of Hund's metals (which however does not apply to Sr$_2$RuO$_4$ in the Fermi liquid regime).

\section{Comparison to experiments}
\label{sec:experiments}
The pairing mechanism and symmetry of the superconducting order parameter of \stront are still outstanding open questions, despite 27 years of intense experimental and theoretical investigations \cite{A.P.MackenzieT.Scaffidi1965a}. New perspectives have recently entirely transformed this field, with the discovery by Pustogow et al.~\cite{Pustogow2019b}  (see also \cite{Ishida2020}) 
that the Knight shift actually sharply drops upon crossing $T_c$, hence challenging the triplet pairing (odd parity) interpretation. For recent discussions and reviews, see e.g. Refs.~\cite{Mackenzie2020, Kivelson2020, Menke2020, Grinenko2020, Sharma2020}. 

In this section, we critically examine whether our theoretical findings are consistent with available experimental evidence, focusing on the two main challengers: the inter-orbital singlet and the $f$-wave singlet (see Tab.~\ref{tab:dmft_leading_gaps}). 
As shown below, we conclude that  the inter-orbital singlet gap is in agreement with basically all experiments, in contrast to the $f$-wave singlet which is ruled out by the momentum distribution of Bogoliubov QPI measurements. 

\subsection{Spin character and two-component gap}
Our two candidate gap functions are both singlets. This is in agreement with the recent NMR finding that the Knight shift drops upon crossing $T_c$~\cite{Pustogow2019b, Ishida2020}, which rules out triplet gaps with an out-of-plane $d$-vector. 

They also both correspond to a two-component order parameter, i.e. transforming under a two-dimensional representation with $E_\mathrm{g}$ symmetry. This is in agreement with recent ultrasound measurements which revealed a jump in the $c_{66}$ elastic constant, a finding which is only consistent with a two-dimensional representation~\cite{Benhabib2020,Ghosh2020} (see, however, ~\cite{Kivelson2020} for an alternative proposal involving two near-degenerate components rather than a symmetry imposed degeneracy). Muon spin-resonance ($\mu$SR) experiments also point at a two-component order parameter~\cite{Grinenko2020}, and the most recent $\mu$SR experiments under hydrostatic pressure further suggest a symmetry imposed degeneracy~\cite{Grinenko2021}.

\subsection{Time-Reversal Symmetry}
\label{sec:TRS}
From our two-component gap functions $(\Delta^{i.o.}_{xy-xz},\Delta^{i.o.}_{xy-yz})$ and $(\Delta^{f}_{xz},\Delta^{f}_{yz})$, we can form the linear combination $\Delta_{\mathrm{TRSI}} = \Delta_1 + \Delta_2$ (or equivalently $\Delta_1-\Delta_2$) which preserves time-reversal invariance (TRSI). It transforms into itself under time-reversal, up to a global sign (Tab.~\ref{tab:dmft_leading_gaps}). In contrast $\Delta_{\mathrm{TRSB}} = \Delta_1 + i \Delta_2$ transforms into its complex conjugate and hence breaks time-reversal symmetry (TRSB). This is also the case of any linear combination $\Delta_1+e^{i\varphi}\Delta_2$ with $\varphi\neq 0, \pi$).

The current experimental situation about time-reversal symmetry is not entirely clear. Kerr effect measurements~\cite{Xia2006}, as well as $\mu$SR~\cite{Luke1998, Luke2000, Grinenko2020, Grinenko2021} suggest TRSB. In contrast, the expected spontaneous magnetization of a TRSB state could not be observed by scanning superconducting quantum interference device microscopy~\cite{Kirtley2007, Hicks2010} or scanning Hall probes~\cite{Curran2014}.
Additionally, recent Josephson tunneling measurements under an applied magnetic field suggest that time-reversal symmetry is preserved (TRSI)~\cite{Kashiwaya2019}. We therefore consider both possibilities in the following, corresponding to the two linear combinations above.

\subsection{Gap Zeros and Lines of Gapless Quasiparticles}
There is strong experimental evidence that \stront hosts gapless excitations, as indicated in particular by measurements of the specific heat~\cite{NishiZaki1999, Nishizaki2000, Deguchi2004, Deguchi2004a, Kittaka2018}, ultrasound attenuation~\cite{Lupien2001}, penetration depth~\cite{Bonalde2000}, directional thermal conductivity~\cite{Hassinger2017} or Bogoliubov QPI~\cite{Sharma2020}.
Furthermore, thermal conductivity and QPI indicate that the gapless quasiparticles reside on lines in momentum space which  run along the c-axis, often referred to as `vertical line nodes'. 
It is however often emphasized in this context that the indications for a two-component order parameter appear to be at odds with `vertical line nodes'. Indeed, the most commonly discussed two-component order parameter has $d_{xz}\propto k_xk_z$, $d_{yz}\propto k_yk_z$ so that the modulus of the order parameter vanishes along `horizontal line nodes' (lying in planes perpendicular to the c-axis). 
As shown below, we find, remarkably, that the inter-orbital singlet state reconciles a two-component order parameter with lines of gapless quasiparticles running along the c-axis. 

\begin{figure*}[!htbp]
  \includegraphics[width=\textwidth]{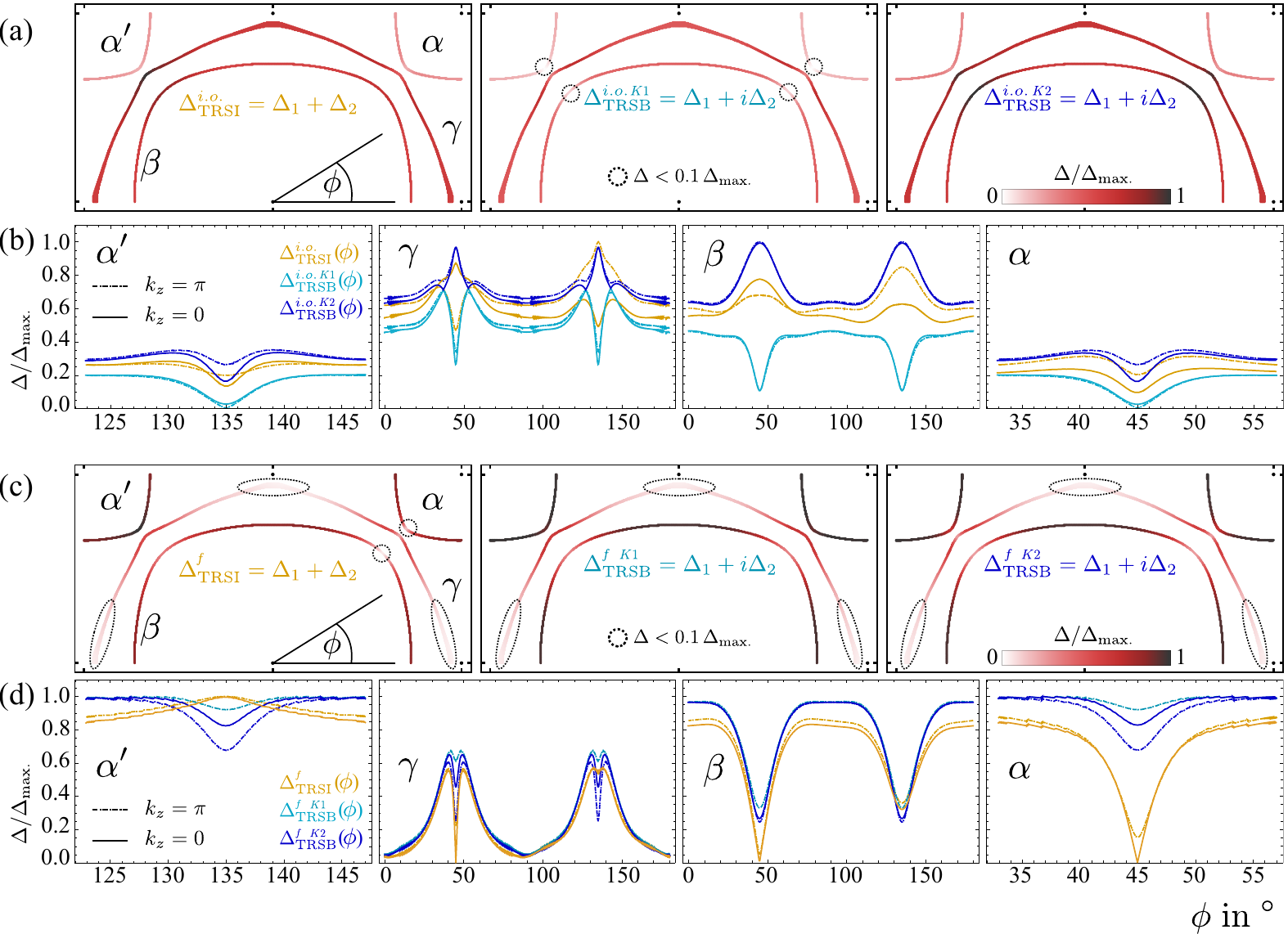}\\
  \caption{ \label{fig:gap_in_band_basis} Superconducting gap function in the eigenbasis of the quasiparticle bands at $\omega=0$. 
  (a) Colormap overlays on the normal state Fermi surface of the inter-orbital gap function $\Delta^{i.o.}$, in the TRSI (left) 
  and the Kramers-split TRSB (center and right) combinations; (b) Angular plots of $\Delta^{i.o.}$ along the four shown Fermi surface sheets as a function of the angle $\Phi$; (c) same as (a) for $\Delta^f$; (d) same as (b) for $\Delta^f$. In panels (a) and (c) the dashed circles/ellipses indicate the location of the gap minima ($\Delta/\Delta_{\mathrm{max}}<0.1$).
  }
\end{figure*}

\begin{figure*}[!htb]
  \includegraphics[width=0.9\textwidth]{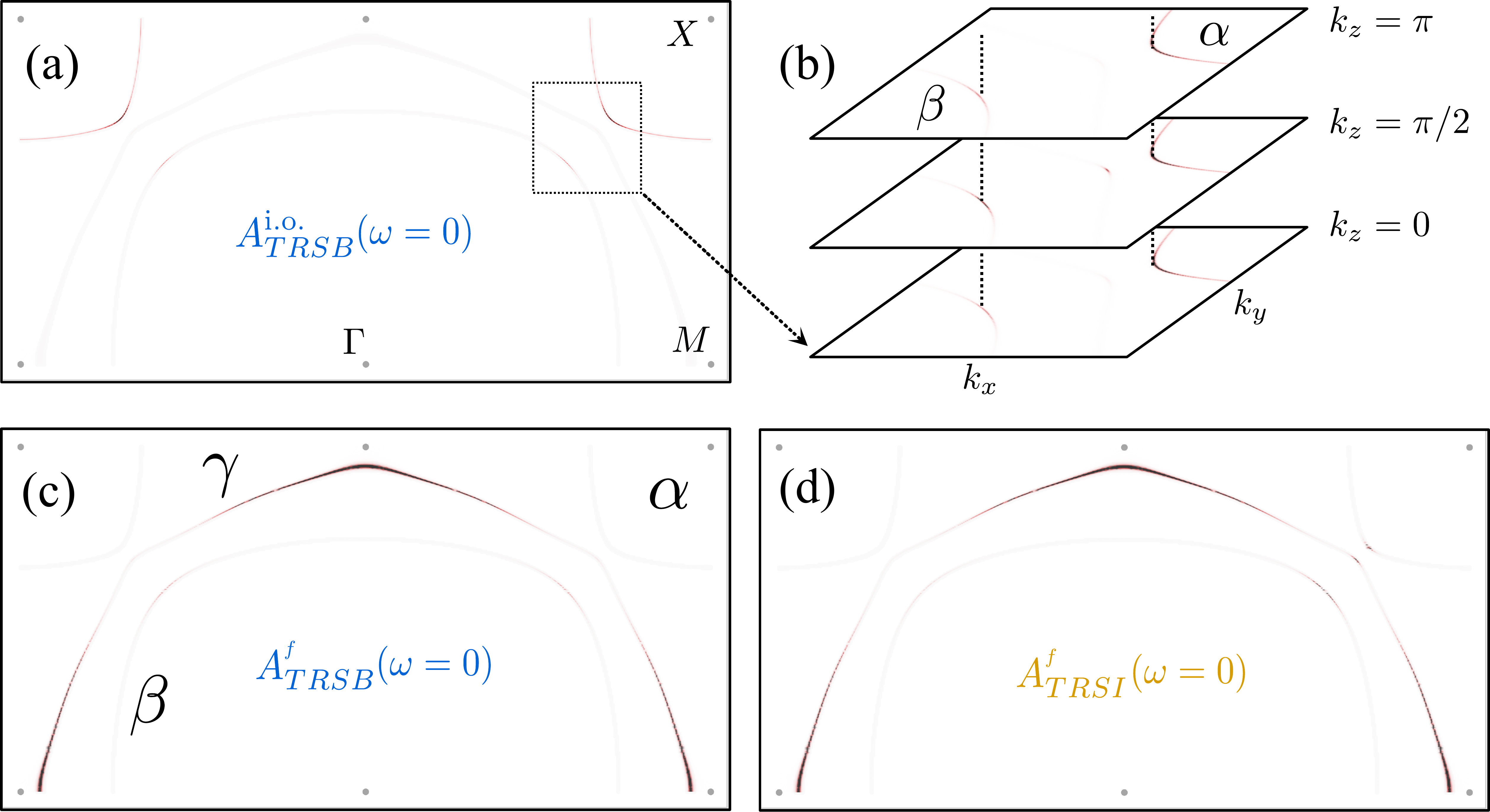}\\
  \caption{\label{fig:resSpec} Spectral weight at $\omega=0$ for a finite gap. (a) $k_z=0$ plane for finite inter-orbital singlet $\Delta^{i.o.}_\text{TRSB}$; (b) $k_z$-dependence of the spectral weight for $\Delta^{i.o.}_\text{TRSB}$; (c)  $k_z=0$ plane for finite $f$-wave singlet $\Delta^{f}_\text{TRSB}$; (d)  $k_z=0$ plane for finite $f$-wave singlet $\Delta^{f}_\text{TRSI}$; as it is fully gapped, the TRSI combination of the inter-orbital singelt is ommited}

\end{figure*}

\subsubsection{Momentum dependence of the gap}
We first consider the momentum-dependence of the gap function. In the previous sections, the gap function was expressed in the orbital basis as $\Delta_{ab}(\mathbf{k})$. Here, we instead consider it in the band basis $\nu \in \{\alpha,\beta,\gamma\}$, performing a basis change:
\begin{equation}
  \label{eq:gapbandbasis}
\Delta_{\mu\nu}(\mathbf{k})\,=\,
\sum_{ab} \langle\psi_{\mathbf{k}\mu}|\varphi_{a}\rangle \Delta_{ab}(\mathbf{k}) \langle\varphi_{b}|\psi_{\mathbf{k}\nu}\rangle 
\end{equation}
where $\langle\psi_{\mathbf{k}\mu}|\varphi_{a}\rangle$ are the overlap matrix elements between
the local orbital states $|\varphi_{a}\rangle$ with $a\in \{xy,xz,yz\}$ and the quasiparticle Bloch states $|\psi_{\mathbf{k}\mu}\rangle$. The interplay of hopping and SOC leads to a strongly mixed $\mathbf{k}$-dependent orbital character of $|\psi_{\mathbf{k}\mu}\rangle$ as can also be seen in Fig.~\ref{fig:FSorb}, where the color code of the Fermi surface sheet corresponds directly to the squared matrix elements $\langle\psi_{\mathbf{k}\mu}|\varphi_{a}\rangle^2$. It is important to note that a strong $\mathbf{k}$-dependence (angular dependence) can be inherited from the matrix elements entering Eq.~\eqref{eq:gapbandbasis}, even when $\Delta_{ab}(\mathbf{k})\approx \Delta_{ab}$ is approximately $\mathbf{k}$-independent in the orbital basis. This is indeed what happens in the case of the inter-orbital pairing.

We plot the normalized absolute value of the resulting gap function in Fig.~\ref{fig:gap_in_band_basis} for TRSB and TRSI combinations of our two component inter-orbital singlet (panels (a) and (b)) and the f-wave singlet (panels (c) and (d)) gaps. We combine color-map overlays on the normal state Fermi surface together with line plots $\Delta(\phi)$ along the four Fermi surface sheets found in the upper half plane of the first BZ for $k_z=0$. The data in Fig.~\ref{fig:gap_in_band_basis} shows that generally linear combinations of the two gap components i) break the C4 symmetry along the $z$-axis of the tetragonal normal state model (due to this we distinguish in our plots between the $\alpha'$ and $\alpha$ sheet) and ii) in the TRSB case break the Kramers degeneracy of the Cooper pair. For the TRSI gaps (orange), we have only one Kramers degenerate line while for the TRSB combinations (light- and dark blue) we show the two split values of the gap (distinguished with superscripts $\Delta^{\text{K1/K2}}$).\\

For the inter-orbital singlet gap function (see panels (a) and (b) of Fig.~\ref{fig:gap_in_band_basis}) the first observation is its overall strong $\mathbf{k}$-dependence which originates almost entirely from the matrix elements $\langle\psi_{\mathbf{k}\mu}|\varphi_{a}\rangle$. The inter-orbital TRSI combination $\Delta^{i.o.}_\text{TRSI}=\Delta^{i.o.}_{xy-xz}+\Delta^{i.o.}_{xy-yz}$ (orange) is finite on \emph{all} Fermi surface sheets. This is due to the SOC driven mixing of orbital character on the sheets. Indeed the lowest absolute value of $\Delta^{i.o.}_\text{TRSI}$ is found on the $\alpha'$ and $\alpha$ sheets (and specifically their intersection with the BZ diagonal) where the admixture of $xy$-character is minimal (see Fig.~\ref{fig:FSorb})\footnote{The finite value of $xy$-character at the $\alpha$-sheets intersection originates from non-local hybridization of $xz$ and $yz$ orbitals. If this hybridization would be neglected, the $xy$-character and indeed the value of the inter-orbital singlet gap would be zero at that point}. Moreover, we observe a pronounced broken $C_4$ symmetry of $\Delta^{i.o.}_\text{TRSI}$. For the specific TRSB combination $\Delta^{i.o.}_\text{TRSB}=\Delta^{i.o.}_{xy-xz}+i\Delta^{i.o.}_{xy-yz}$ (light/dark blue) on the other hand, the $C_4$ symmetry is restored while Kramers degeneracy is broken $\Delta^{i.o \; \text{K1}}_\text{TRSB}\neq\Delta^{i.o \; \text{K2}}_\text{TRSB}$. We find an even more pronounced $\mathbf{k}$-dependence than for the TRSI case and, most remarkably, observe deep minima for $\Delta^{i.o.}_\text{TRSB}$ particularly at the intersection of the $\alpha$, and $\beta$ sheets with the BZ diagonal (marked by dashed black circles). Moreover, we note that there is a slight $k_z$-dependence of the gap which originates from the change in orbital composition of the FS sheets (see dashed lines in panel (b) of Fig.~\ref{fig:gap_in_band_basis}). As can be seen from the plot, however, the deepest gap minima always remain on the $\alpha$ and $\beta$ sheets parallel to the $k_z$-axis  which leads to one dimensional lines of quasiparticle excitations perpendicular to the $k_x$/$k_y$ plane (`vertical line nodes').\\

We now turn to the $f$-wave singlet for which we show data in panels (c) and (d) of Fig.~\ref{fig:gap_in_band_basis}. Here the pronounced $\mathbf{k}$-dependence of the gap in the orbital basis is entangled with that of the projection matrix elements. Also here we find the TRSI combination $\Delta^{f}_\text{TRSI}=\Delta^{f}_{xz}+\Delta^{f}_{yz}$ (orange) to break the $C_4$ symmetry and the specific TRSB combination $\Delta^{f}_\text{TRSB}=\Delta^{f}_{xz}+i\Delta^{f}_\text{yz}$ (light/dark blue) to restore it. Different from the inter-orbital singlet, however, the deepest minima for the $f$-wave singlet occur i) for the TRSI combination and ii) most pronounced on the $\gamma$-sheet where it cuts the $k_x$ and $k_y$ axes where the $xy$-character is the strongest (which is not surprising given that this orbitally diagonal gap is weakest for the $xy$-channel). Also for the $f$-wave singlet gap the $k_z$-dependence is not too strong (see dashed lines in panel (d) of Fig.~\ref{fig:gap_in_band_basis}), so that also here one dimensional quasiparticle lines along $k_z$ occur.

\subsubsection{Gapless Quasiparticles and Momentum-resolved Single-Particle Spectra}
\label{sssec:gapless}
In order to verify that the deep minima of the gap discussed in the previous section lead to quasiparticle excitations in the superconducting phase, we now compute the single particle spectral function for a finite anomalous self-energy. As our results for the anomalous self-energy were performed above $T_c$, we estimate its overall amplitude to match experimental data for the specific heat. To this end we computed the single particle spectral function from the retarded normal propagator of the generalized Green's function Eq.~\eqref{eq:general_G}
\begin{align}
  A(\omega, \mathbf{k}) = -\lim_{\delta \to 0} \frac{1}{\pi} \text{Im}\left[\text{Tr}\left(\mathbf{G}(i\nu_n\rightarrow (\omega+i\delta), \mathbf{k})\right)\right]\,.
\label{eq:spectralfunctionSC}
\end{align}

Using the $\mathbf{k}$-integrated spectral function $A(\omega)$ we then computed the specific heat via \cite{Sigrist2005}
\begin{align}
C_\mathrm{e}(T) = \int_0^\infty d\omega A(\omega) \frac{\omega^2}{k_\mathrm{B}T^2} 
\frac{1}{4\cosh^2{(\omega/2k_\mathrm{B}T)}}\,,
\label{eq:specific_heat}
\end{align}
and found by comparison to experimental data~\cite{Nishizaki2000} amplitudes of $\Delta_\mathrm{max}=0.2/0.35\,\mathrm{meV}$ for the inter-orbital/$f$-wave gap, see App.~\ref{app:specific_heat}. For this procedure we assumed i) a temperature independent anomalous self-energy and ii) fitted its dependence on the fermionic Matsubara frequencies with an analytic model that allowed us to perform the analytic continuation as in Eq.~\eqref{eq:spectralfunctionSC}. Details are reported in App.~\ref{appendix:freq_dep_gaps}.\\

Computation of $A(\omega, \mathbf{k})$ for small energies also allows us to plot the momentum resolved residual spectral weight at $\omega=0$ in the superconducting phase. The resulting plots for the TRSB combination of the inter-orbital pairing and the TRSB and TRSI combinations of the $f$-wave pairing are shown (at $k_z=0$) in Fig.~\ref{fig:resSpec}. In the plots (for which we have chosen a broadening factor in energy of $\delta=10\,\mathrm{meV}$) we overlay the spectral weight (red-color) on a gray background of the normal state Fermi surface. Close inspection shows that the results are in agreement with our expectations from the plots of $\Delta$ in the band basis.
As in the previous section, we start our discussion with the inter-orbital singlet pairing. The TRSI combination $(\Delta^{i.o.}_{xy-xz}+\Delta^{i.o.}_{xy-yz})$ (not shown on Fig.~\ref{fig:resSpec}) leads to a fully gapped spectrum showing no residual spectral weight at the chosen energy resolution. The TRSB combination $(\Delta^{i.o.}_{xy-xz}+i\Delta^{i.o.}_{xy-yz})$, on the other hand, produces sizable weight in the regions of the gap minima on the $\alpha$ and $\beta$ sheets (highlighted by dashed circles) located along the $|k_x|=|k_y|$ diagonals. Further we point out an interesting effect for the inter-orbital pairing: the quasiparticle poles in the inter-orbital superconducting phase are not precisely found at the same momenta as the normal state Fermi surface. Indeed such momentum shifts of quasiparticles is generally expected for inter-orbital anomalous self-energies and was pointed out before \cite{Kaba2019b}. In App.~\ref{app:IOgap} we demonstrate this for a simple two-band model.
We now turn to the $f$-wave singlet pairing. Also here we find the anticipated distribution of spectral weight. While there is some weight on the $\alpha$ and $\beta$ sheets along the zone diagonal, most of the spectrum is found in the vicinity of the broad gap minima at the intersection of the $\gamma$ pocket with the $k_x=0$ and $k_y=0$ lines.

The results we show in Fig.~\ref{fig:resSpec} can be directly compared to the symmetry analysis of Bogoliubov QPI experiments in \cite{Sharma2020}. This comparison likely rules out the $f$-wave singlet pairing as the experimentalists conclude for nodes/minima of the gap function in close vicinity of the normal state $\alpha$ and the $\beta$ pockets along the diagonals. The only candidate which is in satisfactory agreement with this conclusion is the inter-orbital TRSB pairing $(\Delta^{i.o.}_{xy-xz}+i\Delta^{i.o.}_{xy-yz})$. We also point out that the Bogoliubov QPI experiments suggest an approximately fulfilled $C_4$ symmetry of the QPI pattern. It should be noted though, that the slight deviations from full $C_4$ symmetry in favor of a lower $C_2$ axes in the ``unsymmetrized'' raw data provided in the supplemental material of \cite{Sharma2020} could also be explained by a slight deviation from the perfect relative $\pi/2$-phase of the $\Delta^{i.o.}$ components in the TRSB combination.

\section{Conclusion}
\label{sec:conclusion}
We have solved the linearized Eliashberg equation for Sr$_2$RuO$_4$ 
with a DMFT based approximation at intermediate temperatures, and compared the leading symmetry channels of the superconducting order parameters to a large set of experiments. Our main candidate for the superconducting order is 
a two component inter-orbital spin singlet with even spatial parity, and orbital antisymmetry.
The time reversal symmetry breaking combination (TRSB) gap-function has a strong angular dependence with deep minima along the Fermi surface sheets and is compatible with the experiments, including one dimensional lines of quasiparticles parallel to the $c$-axis, located on the $\alpha$ and $\beta$-sheets of the normal state Fermi surface, on the zone diagonal. The driving  force behind the inter-orbital singlet pairing appears to be local inter-orbital spin-correlations. This indicates that Hund's coupling, which already dominates response functions in the normal state, remains key also for the superconducting pairing in \stront.
Finally, the inter-orbital order parameter is remarkably local when considered in the orbital basis. Therefore, it should be possible to obtain it directly within a multiorbital single-site DMFT study at low-temperature inside the superconducting phase, which is certainly an interesting direction for future work. 

\section{Acknowledgements}
We acknowledge discussions with X.~Cao, O.~Gingras and H.~Menke. 
We acknowledge financial support by the DFG project HA7277/3-1.
The Flatiron Institute is a division of the Simons Foundation.

\bibliography{literature}

\appendix


 \section{Generalized propagators}
 \label{appendix:prop_self}
 In the basis of the multi-orbital Nambu spinor Eq.~\eqref{eq:Nambu_spinor} the interacting generalized propagator reads
 \begin{widetext}
   \begin{align}
     \begin{split}
       \hat{\mathbf{G}}_{ab}(\mathbf{k}, \tau)
       &=
       -\left\langle
         \mathcal{T}_\tau
         \mathbf{\Psi}_a(\mathbf{k}, \tau)
         \mathbf{\Psi}_b(\mathbf{k}, 0)^\dagger
       \right\rangle
       \\
       =
       -&\left\langle
       \mathcal{T}_\tau
         \begin{pmatrix}
           c^{}_{a\uparrow}(\mathbf{k}, \tau)c^{\dagger}_{b\uparrow}(\mathbf{k}, 0) & c^{}_{a\uparrow}(\mathbf{k}, \tau)c^{\dagger}_{b\downarrow}(\mathbf{k}, 0) & c^{}_{a\uparrow}(\mathbf{k}, \tau)c^{}_{b\uparrow}(-\mathbf{k}, 0) & c^{}_{a\uparrow}(\mathbf{k}, \tau)c^{}_{b\downarrow}(-\mathbf{k}, 0)\\
           c^{}_{a\downarrow}(\mathbf{k}, \tau)c^{\dagger}_{b\uparrow}(\mathbf{k}, 0) & c^{}_{a\downarrow}(\mathbf{k}, \tau)c^{\dagger}_{b\downarrow}(\mathbf{k}, 0) & c^{}_{a\downarrow}(\mathbf{k}, \tau)c^{}_{b\uparrow}(-\mathbf{k}, 0) & c^{}_{a\downarrow}(\mathbf{k}, \tau)c^{}_{b\downarrow}(-\mathbf{k}, 0) \\
           c^{\dagger}_{a\uparrow}(-\mathbf{k}, \tau)c^{\dagger}_{b\uparrow}(\mathbf{k}, 0)& c^{\dagger}_{a\uparrow}(-\mathbf{k}, \tau)c^{\dagger}_{b\downarrow}(\mathbf{k}, 0)& c^{\dagger}_{a\uparrow}(-\mathbf{k}, \tau)c^{}_{b\uparrow}(-\mathbf{k}, 0)& c^{\dagger}_{a\uparrow}(-\mathbf{k}, \tau)c^{}_{b\downarrow}(-\mathbf{k}, 0) \\
           c^{\dagger}_{a\downarrow}(-\mathbf{k}, \tau)c^{\dagger}_{b\uparrow}(\mathbf{k}, 0) & c^{\dagger}_{a\downarrow}(-\mathbf{k}, \tau)c^{\dagger}_{b\downarrow}(\mathbf{k}, 0) & c^{\dagger}_{a\downarrow}(-\mathbf{k}, \tau)c^{}_{b\uparrow}(-\mathbf{k}, 0) & c^{\dagger}_{a\downarrow}(-\mathbf{k}, \tau)c^{}_{b\downarrow}(-\mathbf{k}, 0) \\
         \end{pmatrix}
       \right\rangle
       \,,
     \end{split}
     \label{eq:general_GF_basis_expression}
   \end{align}
 \end{widetext}
 with orbital indices $a$, $b$ and time-ordering operator $\mathcal{T}_\tau$.
 
\section{Inter-orbital gap functions}
\label{app:IOgap}
In order to illustrate how inter-orbital gap functions can shift spectral weight in momentum space at and around the Fermi energy $\varepsilon_F$, we consider this effect in the simplest possible model of two non-hybridized orbitals. We assume an inter-orbital singlet anomalous self-energy
\begin{align}
	\Delta^{\sigma\sigma'}_{ab}=\delta(\sigma-\sigma')\delta(1-|a-b|) \Delta\,,
\end{align}
where $a,\,b\in\{1,2\}$ are orbital indices and $\sigma,\sigma' \in \{\uparrow, \downarrow\}$ are spin indices.

The generalized Green's function on the Nambu spinor basis Eq.~\eqref{eq:Nambu_spinor} reads
\begin{widetext}
  \begin{align}
    \begin{split}
      \hat{\mathbf{G}}(\mathbf{k}, i\omega_n)=&\\
      \begin{pmatrix}
        i\omega_n-\varepsilon_{1}^{\uparrow}(\mathbf{k}) & 0 & 0 & 0 & 0 & 0 & 0 & \Delta \\
        0 & i\omega_n-\varepsilon_{2}^{\uparrow}(\mathbf{k}) & 0 & 0 & 0 & 0 & -\Delta & 0 \\
        0 & 0 & i\omega_n-\varepsilon_{1}^{\downarrow}(\mathbf{k}) & 0 & 0 & \Delta& 0 & 0 \\
        0 & 0 & 0 & i\omega_n-\varepsilon_{2}^{\downarrow}(\mathbf{k}) & -\Delta & 0 & 0 & 0 \\
        0 & 0 & 0 & \Delta^* & i\omega_n+\varepsilon_{1}^{\uparrow}(\mathbf{k}) & 0 & 0 & 0 \\
        0 & 0 & -\Delta^* & 0 & 0 & i\omega_n+\varepsilon_{2}^{\uparrow}(\mathbf{k}) & 0 & 0 \\
        0 & \Delta^* & 0 & 0 & 0 & 0 & i\omega_n+\varepsilon_{1}^{\downarrow}(\mathbf{k}) & 0 \\
        -\Delta^* & 0 & 0 & 0 & 0 & 0 & 0 & i\omega_n+\varepsilon_{2}^{\downarrow}(\mathbf{k}) \\
      \end{pmatrix}^{-1}
    \end{split}
    \label{eq:twobandmodel}
  \end{align}
 \end{widetext}
where $\varepsilon_{a}^{\sigma}(\mathbf{k})$ is the single particle dispersion of orbital $a$ with spin $\sigma$. We can solve Eq.~\eqref{eq:twobandmodel} and extract the spectral function (for finite $\Delta$) as the trace of the imaginary part of the normal retarded Green's function
\begin{align}
 A(\omega, \mathbf{k}) = -\lim_{\delta \to 0} \frac{1}{\pi} \text{Im}\left[\text{Tr}\left(\mathbf{G}(i\nu_n\rightarrow (\omega+i\delta), \mathbf{k})\right)\right]\,.
\end{align}
Assuming $\varepsilon_{a}^{\uparrow}(\mathbf{k})=\varepsilon_{a}^{\downarrow}(\mathbf{k})=\varepsilon_{a}(\mathbf{k})$, we find at $\omega=0$ the expression
\begin{equation}
A(\omega=0,\mathbf{k})=\frac{2 \delta  \left(2 \Delta ^2+\varepsilon_{1}(\mathbf{k})^2+\varepsilon_{2}(\mathbf{k})^2\right)}{\left(\Delta ^2+\varepsilon_{1}(\mathbf{k}) \varepsilon_{2}(\mathbf{k})\right)^2}\bigg\rvert_{\delta\rightarrow 0}
\label{eq:FSoffDiag}
\end{equation}
which has poles along the contour
\begin{equation}
  \Delta^2=-\varepsilon_{1}(\mathbf{k}) \varepsilon_{2}(\mathbf{k})\, .
  \label{eq:contour}
\end{equation}
This is in stark contrast to an orbital diagonal gap which would be fully gapped: 
\begin{equation}
  \overline{A}(\omega=0,\mathbf{k})=\delta  \left(\frac{2}{\Delta_\text{diag.} ^2+\varepsilon_{1}(\mathbf{k})^2}+\frac{2}{\Delta_\text{diag.}^2+\varepsilon_{2}(\mathbf{k})^2}\right)\bigg\rvert_{\delta\rightarrow 0} \;.
\end{equation}

From Eqs.~\eqref{eq:FSoffDiag} and~\eqref{eq:contour} we see directly that even an inter-orbital $\mathbf{k}$-independent anomalous self-energy leads to a $\mathbf{k}$-dependent gap in the single-particle spectrum. Even more important is the realization that the quasiparticle poles that form the Bogoliubov Fermi surface in $\mathbf{k}$-space no longer coincide with the Fermi surfaces of the normal state (see also \cite{Brydon2018}). Different from an orbital diagonal pairing it is therefore \emph{impossible} to extract the Bogoliubov Fermi surface by finding the intersections of the normal state Fermi surface and the zeros of $\Delta$.

Therefore, in order to address the topic of one dimensional gapless nodal quasiparticles along ``line nodes'' (which are found roughly along $k_z$ in Sr$_2$RuO$_4$) we always need to explicitly calculate the poles of the single-particle propagator. In our two-orbital toy model the contour \eqref{eq:contour} still leads to two dimensional Fermi surface sheets in the three dimensional BZ. In the three orbital model, however, the interplay of spin-orbit-coupling and inter-orbital $\Delta$ can lead to one dimensional quasiparticle lines\cite{Kaba2019b} as is the case for the inter-orbital prime candidate in \stront.

\section{Dependence on the number of bosonic Matsubara frequencies}
\label{appendix:multiple_bosonic}
\begin{figure}[!htb]
	\includegraphics{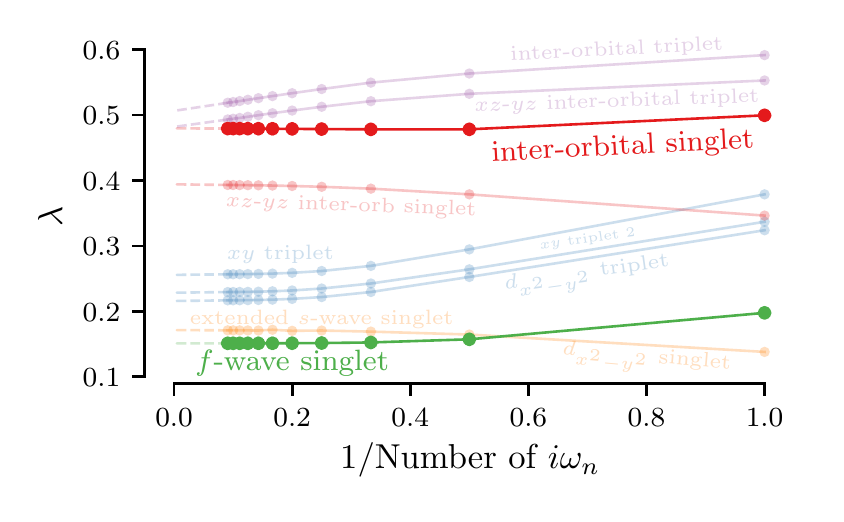}  
	\caption{Using more than one bosonic frequency for the DMFT vertex does not change the relative dominance for our gaps significantly and even makes the inter-orbital singlet more dominant.}
	\label{fig:multiple_bosonic}  
\end{figure}

To investigate if our results for the sampled two particle Green's function $G^{(2)}$ are robust w.r.t. the number of considered bosonic Matsubara frequencies we calculate explicitly the dependencies of the Eliashberg eigenvalues $\lambda$ from one up to eleven bosonic frequencies at $T\approx386\,\mathrm{K}$ and then extrapolate to an infinite number of frequencies, see Fig. \ref{fig:multiple_bosonic}. The data shows that i) the $\lambda$ values of our prime candidate gaps $\Delta^{i.o.}$ and $\Delta^f$ are well converged at the considered numbers of frequencies and ii) that no other gap function trends strongly towards higher values. If these trends do not change to strongly with temperature, our conclusions should be insensitive to increasing the number of bosonic Matsubara frequencies further.

\section{Dependence of the gaps on fermionic Matsubara frequencies}
\label{appendix:freq_dep_gaps}
\begin{figure}[!htb]
	\includegraphics{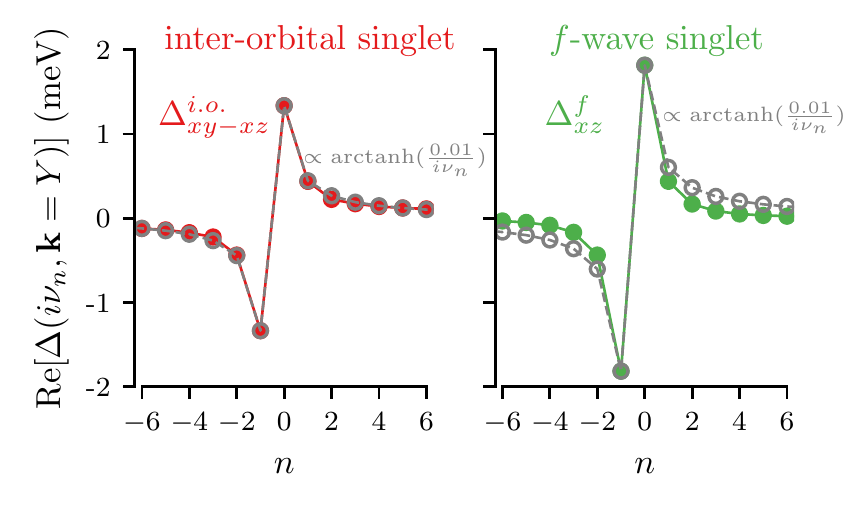}   
	\caption{The inter-orbital singlet and $f$-wave singlet show a frequency dependence which is in a reasonable agreement with the analytic dependence $\propto\mathrm{arctanh}(0.01/i\nu_n)$ behavior. These gaps stem from a calculation with a particle-particle vertex of eleven bosonic Matsubara frequencies.}
	\label{fig:deltas_over_freq}
\end{figure}

From the calculation with eleven bosonic frequencies (App. \ref{appendix:multiple_bosonic}) we also get insight in the fermionic frequency dependence of our gap candidates, see Fig. \ref{fig:deltas_over_freq}. For a calculation with one bosonic frequency the gaps $\Delta(i\nu_n)$ are only nonzero at $i\nu_{n=-1,0}$, however, by increasing the number of bosonic frequencies they become nonzero at higher $i\nu_n$. By using eleven bosonic frequencies they show a converging behavior in reasonable agreement with the analytic frequency dependence $\propto\mathrm{arctanh}(0.01/i\nu_n)$.\\

To motivate this analytic frequency dependence we assume that $\Delta(i\nu_n)$ is analytical and converges for $|i\nu_n|\rightarrow\infty$, which allows us to use the spectral representation
\begin{align}
	\Delta(i\nu_n) = \int_{-\infty}^{\infty} d\omega \frac{A_{\Delta}(\omega)}
	{i\nu_n - \omega}\,.
\end{align}
We consider an imaginary even box-shaped spectral function, 
\begin{align}
	A_{\Delta}(\omega) = i \big[\Theta(\omega+a) - \Theta(\omega-a)\big]\,,
\end{align}
where $2a$ is the box size. This yields 
\begin{align}
\begin{split}
\Delta(i\nu_n)
=\int_{-a}^{a} 
d\omega \frac{i}{i\nu_n - \omega}
= 2 i\mathrm{Arctanh}\left(\frac{a}{i\nu_n}\right)\,,
\end{split}
\label{eq:analytic_matsubara_frequency_dependence}
\end{align}
which is odd in $i\nu_n$ and purely real. Using Eq. \eqref{eq:analytic_matsubara_frequency_dependence} with $a=0.01$ shows a reasonable agreement with the numerical data, see Fig. \ref{fig:deltas_over_freq}, supporting the chosen spectral function.

Note, that by analytical continuing $i\nu_n\rightarrow \omega + i\delta$ Eq. \eqref{eq:analytic_matsubara_frequency_dependence}, as we do in Sec. \ref{sssec:gapless}, we get real frequency gaps $\Delta(\omega)$, which have a finite constant real part and a linear imaginary part. Therefore these gaps do not vanish at $\omega=0\,\mathrm{eV}$ and gap the Fermi surface. This is different to other odd Matsubara frequency gaps, e.g. a linear behavior $\propto i\nu_n$, which leaves the Fermi surface ungapped \cite{Balatsky1992, Linder2019}.

\section{Fitting gap size to specific heat}
\label{app:specific_heat}
In order to estimate realistic overall prefactors of our gap function for the computation of spectral functions, we computed the low temperature specific heat Eq. \eqref{eq:specific_heat} and compared to experimental data~\cite{Nishizaki2000}.

\begin{figure}[!htbp]
	\includegraphics{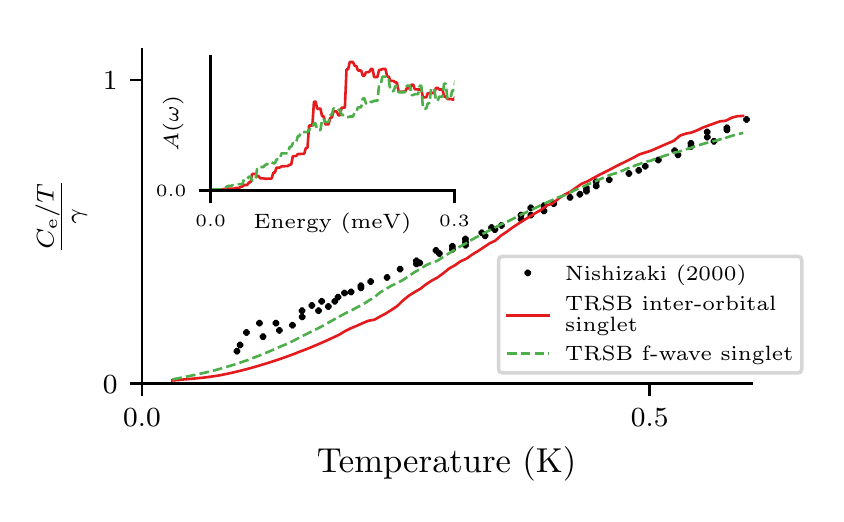}	\label{fig:spectral_function_TRSI}
	\caption{Comparison of experimental data~\cite{Nishizaki2000} and calculated temperature dependence of the specific heat for the inter-orbital singlet and $f$-wave gap (shown here in TRSB combination). The inset shows the $\mathbf{k}$-integrated spectral function up to $0.3$ meV.}
	\label{fig:spectral_function}
\end{figure}

For the comparison, which is shown in Fig.~\ref{fig:spectral_function}, the spectral function was calculated for a $300 \times 300 \times 300$ $\mathbf{k}$-mesh and for $200$ $\omega$-points from $0$ to $1\,\mathrm{meV}$. The integral Eq. \eqref{eq:specific_heat} was solved numerically with SciPy \cite{scipy}.

We calculated the electronic specific heat for multiple gap sizes $\Delta_{\mathrm{max}}$ of the analytic models to find the best fit to the experimental data~\cite{Nishizaki2000}. The best overall agreement was found for $\Delta_{\mathrm{max}} = 0.2\,\mathrm{meV}$ for the inter-orbital and $\Delta_{\mathrm{max}} = 0.35\,\mathrm{meV}$ for the $f$-wave singlet.


\section{Two-particle Green's function sampling}
\label{appendix:impurity_g2_sampling}

We sample both the single-particle $G_{ab}$ and two-particle $G^{(2)}_{abcd}$ Green's functions using CT-HYB and hybridization function removal \cite{Gull2011}. The tetragonal symmetry of Sr$_2$RuO$_4$ causes the single particle Green's function to be diagonal in the $t_{2\mathrm{g}}$ Wannier orbital space, $G_{ab} = \delta_{ab} G_{aa}$. This symmetry is due to the DMFT self-consistency shared with the impurity hybridization function $\widetilde{\Delta}_{ab} =  \delta_{ab} \widetilde{\Delta}_{aa}$. Hence, all non-zero components of $G_{ab}$ can be sampled by removal of hybridization insertions in the expansion of the partition function. This is, unfortunately, not the case for the two particle Green's function $G^{(2)}_{abcd}$. The spin-flip and pair-hopping terms in the local Kanamori interaction generates nonzero orbital combinations $abcd$ that can not be sampled when the hybridization function is diagonal in orbital space. Hence, in order to measure all components of $G^{(2)}_{abcd}$, we perform a single-particle basis transform of the impurity model to a basis were all $ab$ components of $\widetilde{\Delta}_{ab}$ are nonzero. However, an off-diagonal $\widetilde{\Delta}$ generates a Monte Carlo sign problem, making low temperature calculations unfeasible \cite{Strand2019a}. For the present study we had to restrict ourselves to temperatures $T \gtrsim 290\,$K.


\section{Static approximation of the fully irreducible vertex function}
\label{app:ApproxFullIrrVertex}

In our calculations we approximate the fully irreducible vertex $\Lambda^{\mathrm{s/t}}$ with the static first order term $\widetilde{\Lambda}^{\mathrm{s/t}}$, neglecting forth order corrections,
$\Lambda^{\mathrm{s/t}} \approx \widetilde{\Lambda}^{\mathrm{s/t}}$. The first order fully irreducible vertex $\widetilde{\Lambda}^{\mathrm{s/t}}$ is given by
\begin{align}
 \label{eq:Lambda_stat}
  \widetilde{\Lambda}^{\mathrm{s}}
  \equiv
  -
  \frac{1}{2}U^{\mathrm{d}}
  -
  \frac{3}{2}U^{\mathrm{m}}
  \,,
  \\
  \widetilde{\Lambda}^{\mathrm{t}}
  \equiv
  -
  \frac{1}{2}U^{\mathrm{d}}
  +
  \frac{1}{2}U^{\mathrm{m}}
  \,,
\end{align}
where $U^{\mathrm{d/m}}$ are the static interaction tensors of the rotationally invariant Kanamori interaction
\begin{align}
U^{d}_{abcd} &=
\begin{cases}
U, & \mathrm{if}\;a=b=c=d \\
-U'+2J, & \mathrm{if}\;a=d\neq b=c \\
2U'-J, & \mathrm{if}\;a=b\neq c=d \\
J, & \mathrm{if}\;a=c\neq b=d \\
0, & \mathrm{else}
\end{cases}
\\
U^{m}_{abcd} &=
\begin{cases}
U, & \mathrm{if}\;a=b=c=d \\
U', & \mathrm{if}\;a=d\neq b=c \\
J, & \mathrm{if}\;a=b\neq c=d \\
J, & \mathrm{if}\;a=c\neq b=d \\
0, & \mathrm{else}
\end{cases}
\,.
\label{eq:bare_vertex}
\end{align}
with $U'=U-2J$, and use the same values for the Hubbard interaction $U$ and the Hund's $J$ as in our self consistent DMFT calculations, $U=2.3\,\text{eV}$, $J=0.4\,\text{eV}$.


\section{Static approximation of $\Gamma$ in the equation for the vertex ladder}
\label{app:StaticApproxVertexLadder}
Conceptually, the ladder-vertex $\Phi^{\text{d/m}}$ Eq. \eqref{eq:laddervertex} can be approximated using the DMFT impurity vertex $\Gamma_{\text{DMFT}}^\text{d/m}$ and the DMFT generalized lattice susceptibility $\chi^{\text{d/m}}_\text{DMFT}$ obtained from the Bethe-Salpeter equation, see Eq.~\eqref{eq:bse_approx},
\begin{multline}
  \Phi^{\text{d/m}}\left[\Gamma^\text{d/m},\chi^\text{d/m}\right]
  \\ \approx
  \Phi^{\text{d/m}}\left[\Gamma_\text{DMFT}^\text{d/m},\chi_\text{DMFT}^\text{d/m}\right]
  \equiv
  \Phi^{\text{d/m}}_\text{DMFT}
  \, .
\end{multline}
However, $\Phi^{\text{d/m}}_\text{DMFT}$ can not be used in the Parquet equation, due to stochastic quantum Monte Carlo noise at higher frequencies, introduced through $\Gamma^\text{d/m}_\text{DMFT}$. Therefore we resort to the renormalized static approximation $\Gamma^{\text{d/m}}_\text{DMFT} \approx \overline{U}^{\text{d/m}}$, which leads to
\begin{multline}
  \Phi^{\text{d/m}}\left[\Gamma^\text{d/m},\chi^\text{d/m}\right]
   \approx
  \Phi^{\text{d/m}}\left[\overline{U}^{\text{d/m}}, \chi_\text{DMFT}^\text{d/m}\right]
  \\ = 
  \overline{U}^{\mathrm{d/m}}
  \chi_{\text{DMFT}}^{\text{d/m}}(i\omega_n,\mathbf{q}) \, \overline{U}^{\mathrm{d/m}}
  \equiv
  \widetilde{\Phi}^{\text{d/m}}(i\omega_n,\mathbf{q})
  \,,
  \label{eq:laddervertex_static}
\end{multline}
where $\overline{U}^{\text{d/m}}$ are static re-normalized interaction tensors.

Finally we obtain an approximation for the exact particle-particle vertex
$\Gamma^{\text{s/t}}$ Eqs.~\eqref{eq:gamma_triplet_no_approximation} and \eqref{eq:gamma_singlet_no_approximation}
by combining the two static approximations $\Lambda \approx \widetilde{\Lambda}$
Eq.~\eqref{eq:Lambda_stat} and $\Phi \approx \widetilde{\Phi}$ Eq.~\eqref{eq:laddervertex_static}
\begin{equation}
  \Gamma^{\text{s/t}}\left[ \Lambda^\text{s/t}, \Phi^{\text{d/m}} \right]
  \approx
  \Gamma^{\text{s/t}}\left[ \widetilde{\Lambda}^\text{s/t}, \widetilde{\Phi}^{\text{d/m}} \right]
  \equiv
  \widetilde{\Gamma}^{\text{s/t}}
  \, .
\end{equation}

\begin{figure}[!htb]
	\includegraphics{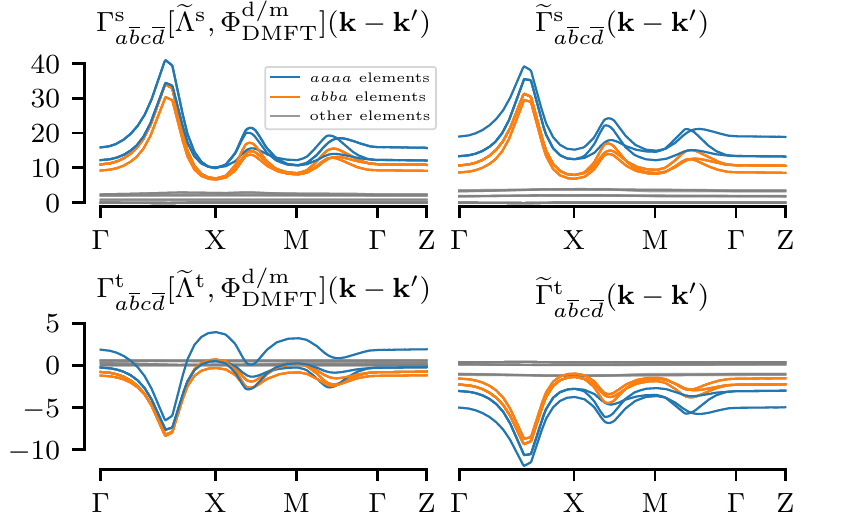}   
	\caption{
          The singlet vertex at the zeroth bosonic Matsubara frequency $\omega_0 = 0$ and first fermionic Matsubara frequencies $\nu_0=\nu'_0=\pi/\beta$ (top left) is well matched by our static interaction approximation (top right), while the triplet vertices differ visibly (bottom).}
	\label{fig:gamma_approx}
\end{figure}

The (orbital dependent) re-normalized interaction tensors $\overline{U}^{\text{d/m}}$ were obtained by minimizing
\begin{equation}
  \label{eq:norm_static_error}
  \min_{\overline{U}^{\text{d/m}}} \sum_{\mathbf{q}}
  \left|  \Gamma^{\text{s/t}}\left[ \widetilde{\Lambda}^\text{s/t}, \Phi_\text{DMFT}^{\text{d/m}} \right]
  -
  \widetilde{\Gamma}^{\text{s/t}}
  \right|^{2}
\end{equation}
in the low frequency limit, i.e.\ for the zeroth bosonic Matsubara frequency $\omega_0 = 0$ and first fermionic Matsubara frequencies $\nu_0=\nu'_0=\pi/\beta$.
In Fig.~\ref{fig:gamma_approx} we show the comparison for $\overline{U}_{xy}=1.2\,\mathrm{eV}$ and $\overline{U}_{xz/yz}=1.1\,\mathrm{eV}$.
The exact structure of $\overline{U}^{\text{d/m}}$ is given by
\begin{align}
\overline{U}^{d}_{abcd} &=
\begin{cases}
\overline{U}_{\mathrm{xy}}, & \mathrm{if}\;a=b=c=d=xy \\
\overline{U}_{\mathrm{xz/yz}}, & \mathrm{if}\;a=b=c=d=xz/yz \\
-\overline{U}'+2\overline{J}, & \mathrm{if}\;a=d\neq b=c \\
2\overline{U}'-\overline{J}, & \mathrm{if}\;a=b\neq c=d \\
\overline{J}, & \mathrm{if}\;a=c\neq b=d \\
0, & \mathrm{else}
\end{cases}
\\
\overline{U}^{m}_{abcd} &=
\begin{cases}
\overline{U}_{\mathrm{xy}}, & \mathrm{if}\;a=b=c=d=xy \\
\overline{U}_{\mathrm{xz/yz}}, & \mathrm{if}\;a=b=c=d=xz/yz \\
\overline{U}', & \mathrm{if}\;a=d\neq b=c \\
\overline{J}, & \mathrm{if}\;a=b\neq c=d \\
\overline{J}, & \mathrm{if}\;a=c\neq b=d \\
0, & \mathrm{else}
\end{cases}
\,.
\label{eq:bare_vertex_re}
\end{align}
Here $\overline{U}_{xy}$ and $\overline{U}_{xz/yz}$ are the free parameters, while the others are constrained by $\overline{J}/\overline{U}_{\mathrm{avg}} = 0.4/ 2.3$ and $\overline{U}' = \overline{U}_{\mathrm{avg}} -2\overline{J}$ with $\overline{U}_{\mathrm{avg}} = 1/3 \cdot(\overline{U}_{xy} + 2\overline{U}_{xz/yz})$.
%
%


\section{Numerical details}
\label{app:NumericalDetails}

We sample the one- and two-particle Green's function with the CT-HYB solver implemented in the TRIQS project \cite{Parcollet2015, Seth2016}. We use a $32\times 32 \times 32$ k-mesh and do $10^9$ ($2.1\times 10^8$) Monte Carlo cycles for the one- (two-) particle Green's function. For sampling the two-particle Green's function $G^{(2)}$ we employ the particle-hole notation using a $40\times40$ fermionic frequency grid and only one bosonic frequency. To investigate if the restriction of using only one bosonic frequency is critical for our results, we additionally sample $G^{(2)}$ for eleven bosonic frequencies at $T\approx386\,\mathrm{K}$, c.f. App.~\ref{appendix:multiple_bosonic}.

We solve the impurity/lattice Bethe-Salpeter equation to obtain $\Gamma^{\mathrm{d/m}}$/$\chi^{\mathrm{d/m}}$~\cite{Strand2019a} with the two-particle response function toolbox (TPRF) of the TRIQS library \cite{TPRF}. Both the impurity and lattice Bethe-Salpeter equations are matrix equations in (fermionic) Matsubara space, which we solve using three finite frequency cut-offs $n_\nu =20, 30, 40$ from which we then extrapolated $\chi^{\mathrm{d/m}}$ to infinite frequencies.

The linearized Eliashberg equation Eq.~\eqref{eq:eliashberg} is technically a large eigenvalue problem in a vector space spanned by the orbital, frequency, and momentum indices. To find (generally complex) eigenvectors $\Delta$ and their corresponding eigenvalues $\lambda$ we employ TPRF~\cite{TPRF}. In TPRF the implicit matrix-vector product
$(\Gamma G G) \cdot \Delta$ is implemented using fast Fourier transforms, and
the highest eigenvalues $\lambda$ and eigenvectors $\Delta$ are determined using the implicitly restarted Arnoldi method from the ARPACK library~\cite{ARPACK} through the SciPy package~\cite{scipy}. The Pauli principle (i.e. the $SPOT$-condition Eq.~\eqref{eq:SPOT_def}) is exploited to constrain the solutions to all possible allowed symmetries, further reducing the computational effort. No other restrictions are imposed on $\Delta$. 

\section{Scaling $\chi^{\mathrm{d/m}}$ for channel analysis}
\label{appendix:scaling_chi}

We distinguish between magnetic-, density-, inter-/intra-orbital and local fluctuations in our channel analysis  by scaling selected components of $\chi^{\mathrm{d/m}}$ before they enter the equation for the pairing vertices Eqs. \eqref{eq:gamma_singlet_no_approximation} and \eqref{eq:gamma_triplet_no_approximation} via the reducible vertex function Eq. \eqref{eq:laddervertex}.
To study the effect of magnetic (density) fluctuations we scale the complete $\chi^{\mathrm{m}}$ ($\chi^{\mathrm{d}}$) tensor.
For the intra-orbital fluctuations we solely scale the components of the density and magnetic susceptibility where all indices are equal, i.e. only the $\chi^{\mathrm{d/m}}_{aaaa}$ components.
Complementary, for the inter-orbital fluctuations we scale all components except for the $\chi^{\mathrm{d/m}}_{aaaa}$ components.
For studying local fluctuations we first calculate the local susceptibilities via
\begin{align}
	\chi^{\mathrm{d/m}}_{\mathrm{loc}}
	=
	\frac{1}{N_{\mathbf{k}}}
	\sum_{\mathrm{BZ}}
	\chi^{\mathrm{d/m}}(\mathbf{k})
	\,,
\end{align}
which allows us to express the susceptibilities as a sum of $\chi^{\mathrm{d/m}}_{\mathrm{loc}}$ and a dispersive part, i.e.
\begin{align}
\chi^{\mathrm{d/m}}(\mathbf{k})
=
\chi^{\mathrm{d/m}}_{\mathrm{local}}
+
\chi^{\mathrm{d/m}}_{\mathrm{dispersive}}(\mathbf{k})
\,.
\end{align}
We then only scale $\chi^{\mathrm{d/m}}_{\mathrm{loc}}$ for the channel analysis.

\section{Random phase approximation}
\label{sec:results_rpa}

\begin{figure}[h]
\includegraphics{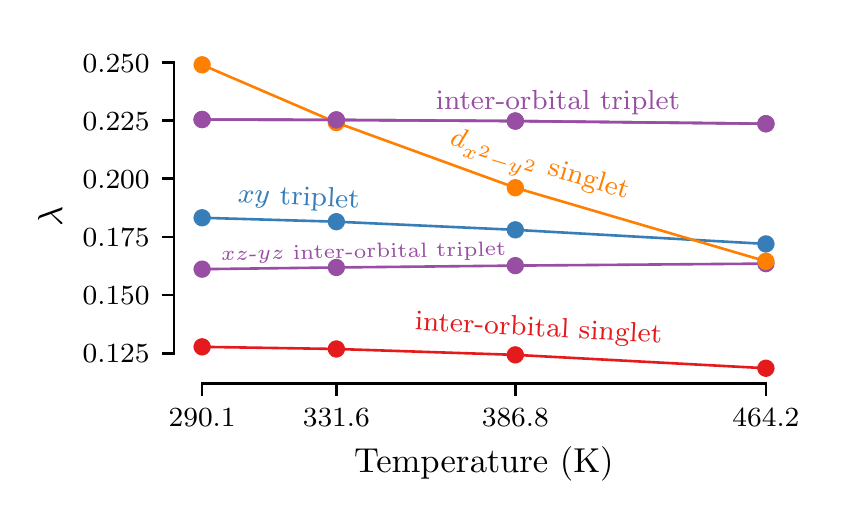}     
\caption{The $d_{x^2-y^2}$ singlet gap takes the lead for lower temperatures in RPA. The doubly degenerate inter-orbital singlet, which is the dominant one in DMFT, remains low tier.}
\label{fig:rpa_lambda_over_temperature}
\end{figure}

As a benchmark we solve the linearized Eliashberg equation in the random phase approximation (RPA). In RPA the interaction is only treated on the level of the two-particle Green's function, hence we use the spin-independent non-interacting Green's function
\begin{align}
  G^{0}_{ab}(\nu_n, \mathbf{k}) = \left[{(i\nu_n + \mu) \mathbf{1} -
  \epsilon_{\mathbf{k}}}\right]^{-1}_{ab}\,,
  \label{eq:non_interacting_greens_function}
\end{align}
where $\mu$ is the chemical potential and $\epsilon_{\mathbf{k}}$ the dispersion relation, instead of $G$ inside Eq. \eqref{eq:eliashberg}. To construct the particle-particle vertices in RPA we use the equations introduced in App.~\ref{app:ApproxFullIrrVertex} and ~\ref{app:StaticApproxVertexLadder}, but use the RPA density/magnetic susceptibilities given by
\begin{align}
  \chi^{\text{d/m, RPA}}_{abcd}(Q) =
  \frac{\chi^{0,\mathrm{d/m}}(Q)}
  {1 \pm U^{\text{d/m}}
  \chi^{0,\mathrm{d/m}}(Q)}
  \,,
  \label{eq:rpa}
\end{align}
with the bare susceptibility 
\begin{align}
  \chi^{0,\mathrm{d/m}}_{abcd}(Q) = -\frac{1}{N_{\mathbf{k}}\beta}\sum_{K}
  G^{0}_{bc}(K) G^{0}_{da}(K+Q)\,,
  \label{eq:bare_susceptibility}
\end{align}
and the local vertex $U^{\text{d/m}}$ given in Eq. \eqref{eq:bare_vertex}.
We use the renormalized interaction parameters $U=0.575\,\text{eV}$ and $J=0.1\,\text{eV}$, which have the same $U/J$ ratio as the DMFT parameters but are smaller, because otherwise the RPA susceptibilities diverge.

We identify the following leading superconducting gap functions: a $d_{x^2-y^2}$ singlet, a doubly degenerate inter-orbital triplet that couples $xy$ with $xz/yz$, an odd-frequency triplet, an inter-orbital triplet that couples $xz$ with $yz$ and the doubly degenerate inter-orbital singlet that couples $xy$ with $xz/yz$, see Tab. \ref{tab:rpa_leading_gaps}. Their eigenvalues over temperature is pictured in Fig. \ref{fig:rpa_lambda_over_temperature}.

While the doubly degenerate inter-orbital triplet leads for the higher temperatures, the $d_{x^2-y^2}$ singlet rapidly increases for lower temperatures where it becomes dominant. This result is in agreement with other RPA studies \cite{Zhang2018i, Romer2019b}, but also with RPA-like schemes that try to capture correlation effects by dressing the non-interacting Green’s function \cite{Gingras2019}. The doubly degenerate inter-orbital singlet, which is dominant in our DMFT calculation, is also present among the leading gaps, but remains low tier.
\begin{table*}
	\begin{threeparttable}
		\renewcommand{\arraystretch}{1.4}
		\caption{The first five leading gaps in descending order in RPA at $T\approx290\,\text{K}$, $U=0.575\,\text{eV}$ and $J=0.1\,\text{eV}$ (c.f. Fig. \ref{fig:rpa_lambda_over_temperature}). Gaps with the same $SPOT$ symmetry share the same color. As many of our gap functions have not only a single finite matrix element in the orbital space the column ``Orbital character'' indicates the dominant orbital matrix element.
	}
		\begin{tabularx}{0.85\textwidth}{ >{\hsize=0.8\hsize}L<{\kern\tabcolsep}@{}
				>{\hsize=0.2\hsize}C<{\kern\tabcolsep}@{}
				>{\hsize=0.2\hsize}C<{\kern\tabcolsep}@{}
				>{\hsize=0.2\hsize}C<{\kern\tabcolsep}@{}
				>{\hsize=0.2\hsize}C<{\kern\tabcolsep}@{}
				>{\hsize=0.8\hsize}L<{\kern\tabcolsep}@{}}
			\hline\hline
			&\multicolumn{4}{c}{Symmetries} &\\[-.45cm]
			\cline{2-5} \\[-1cm]
			Pairing & Spin & Parity & Orbital & Time & \hspace{1.05cm}Orbital character  \\
			\hline
			$d_{x^2-y^2}$ singlet & \textcolor{singlet_d_x2_y2}{$\bm{-}$} & \textcolor{singlet_d_x2_y2}{$\bm{+}$} & \textcolor{singlet_d_x2_y2}{$\bm{+}$} & \textcolor{singlet_d_x2_y2}{$\bm{+}$} &\hspace{2.85cm} intra $xy$  \\
			inter-orbital triplet & \textcolor{inter_orb_triplet_xyxz_xyyz}{$\bm{+}$} & \textcolor{inter_orb_triplet_xyxz_xyyz}{$\bm{+}$} & \textcolor{inter_orb_triplet_xyxz_xyyz}{$\bm{-}$} & \textcolor{inter_orb_triplet_xyxz_xyyz}{$\bm{+}$} &\hspace{1.06cm}degenerate $\begin{cases} \text{inter }xy\text{-}yz \\ \text{inter }xy\text{-}xz \end{cases}$ \\
			$xy$ triplet &\textcolor{triplet_d_x2_y2}{$\bm{+}$} & \textcolor{triplet_d_x2_y2}{$\bm{+}$} & \textcolor{triplet_d_x2_y2}{$\bm{+}$} & \textcolor{triplet_d_x2_y2}{$\bm{-}$} &\hspace{2.85cm} intra $xy$ \\
			$xz$-$yz$ inter-orbital triplet & \textcolor{inter_orb_triplet_xyxz_xyyz}{$\bm{+}$} & \textcolor{inter_orb_triplet_xyxz_xyyz}{$\bm{+}$} & \textcolor{inter_orb_triplet_xyxz_xyyz}{$\bm{-}$} & \textcolor{inter_orb_triplet_xyxz_xyyz}{$\bm{+}$} & \hspace{2.85cm}  inter $xz$-$yz$ \\
			inter-orbital singlet & \textcolor{inter_orb_singlet_xyxz_xyyz}{$\bm{-}$} & \textcolor{inter_orb_singlet_xyxz_xyyz}{$\bm{+}$} & \textcolor{inter_orb_singlet_xyxz_xyyz}{$\bm{-}$} & \textcolor{inter_orb_singlet_xyxz_xyyz}{$\bm{-}$} &\hspace{1.06cm}degenerate $\begin{cases} \text{inter }xy\text{-}yz \\ \text{inter }xy\text{-}xz \end{cases}$ \\
			\hline
			\hline
		\end{tabularx}
		\label{tab:rpa_leading_gaps}
	\end{threeparttable}
\end{table*}
\end{document}